\documentclass[twocolumn]{autartArxiv}    
\usepackage{graphicx}                            
\usepackage{amsmath, amssymb}
\usepackage{breqn}
\usepackage{xcolor}
\usepackage{amsfonts}
\usepackage{mathtools}
\usepackage{comment}

\newtheorem{theorem}{Theorem}[section]
\newtheorem{lemma}{Lemma}[section]
\newtheorem{corollary}{Corollary}[section]
\newtheorem{remark}{Remark}[section]
\newtheorem{proposition}{Proposition}[section]

\newcommand{\bea}{\begin{eqnarray}}
\newcommand{\eea}{\end{eqnarray}}
\newcommand{\sign}{{\rm sign}}
\newcommand{\diag}{{\rm diag}}
\newcommand{\tr}{{\rm tr}}
\newcommand{\T}{{\rm T}}
\newcommand{\R}{\mathbb{R}}

\newcommand{\beq}{\begin{equation}}
\newcommand{\eeq}{\end{equation}}

\DeclareSymbolFont{newfont}{OML}{cmm}{m}{it}
\DeclareMathSymbol{\Epsilon}{3}{newfont}{15}
\DeclareMathSymbol{\Varrho}{3}{newfont}{37}

\allowdisplaybreaks

\begin{document}

\begin{frontmatter}

\title{Partial-State Feedback Multivariable MRAC and Reduced-Order Designs \thanksref{footnoteinfo}} 

\thanks[footnoteinfo]{This paper was not presented at any IFAC 
	meeting. Corresponding author G.~Tao. Tel. +1-434-924-4586. 
	Fax +1-434-924-8818. \\ This work was supported by NSF under grant ECCS1509704.}

\author{Ge Song}\ead{gns5mq@virginia.edu}~and~    
\author{Gang Tao}\ead{gt9s@virginia.edu}               

\address{Department of Electrical and Computer Engineering, University of Virginia \\ Charlottesville, VA 22904, USA}  
          

\begin{abstract}                          
This paper develops a new model reference adaptive control (MRAC) framework using partial-state feedback for solving a multivariable adaptive output tracking problem. The developed MRAC scheme has full capability to deal with plant uncertainties for output tracking and has desired flexibility to combine the advantages of full-state feedback MRAC and output feedback MRAC. With such a new control scheme, the plant-model matching condition is achievable as with an output or state feedback MRAC design. A stable adaptive control scheme is developed based on LDS decomposition of the plant high-frequency gain matrix, which guarantees closed-loop stability and asymptotic output tracking. The proposed partial-state feedback MRAC scheme not
only expands the existing family of MRAC, but also provides new features to the adaptive control system, including additional design flexibility and feedback capacity. Based on its additional design flexibility, a minimal-order MRAC scheme is also presented, which reduces the control adaptation complexity and relaxes the feedback information requirement, compared to the existing MRAC schemes. New results are presented for plant-model matching, error model, adaptive law and stability analysis. A simulation study of a linearized
aircraft model is conducted to demonstrate the effectiveness and new features of the proposed MRAC control scheme.
\end{abstract}

\end{frontmatter}

\section{Introduction}
Adaptive control is an effective control methodology which can deal with system uncertainties. In the past decades, it attracted tremendous attentions \cite{m80}--\cite{d13b}. Recently, more results have been developed such as adaptive backstepping control \cite{zw08b}--\cite{zww09}, adaptive posicast control \cite{dasl13}, adaptive sliding mode control \cite{c06}, robust adaptive control \cite{ot89}--\cite{opn15} and other adaptive control designs \cite{oha03}--\cite{bjj14}. 

Model reference adaptive control (MRAC) is a main and mature branch among various adaptive control techniques. One of its essential features is the capability of ensuring asymptotic output or state tracking of a given reference model system and closed-loop signal boundedness in the presence of system uncertainties \cite{chik03}, \cite{hb12}. During the recent decades, multivariable MRAC theory for multi-input and multi-output systems, such as aircraft systems, has evolved into a mature branch. Considerable effort has been devoted to the development of multivariable MRAC theory. The existing results include (i) state feedback MRAC for state tracking \cite{na89}, \cite{is96}, \cite{ysb16}, \cite{tjm01}; (ii) state feedback MRAC for output tracking \cite{bg97}, \cite{gtl11}; and (iii) output feedback MRAC for output tracking \cite{gs84}--\cite{wtl15}. The controller structure of state feedback MRAC for state tracking is simple, but the plant-model state matching condition is restrictive, which can only be satisfied for system matrices in certain canonical forms. Compared to state feedback MRAC for state tracking, state feedback MRAC for output tracking is suitable for more applications because of its unrestrictive matching condition and simple controller structure \cite{gtl11}, although the full-state vector requirement may still confine its applications. When the full state vector is hard to be obtained, output feedback MRAC for output tracking attracts more attention, although its controller structure is more complex that may limit its applications.

\textbf{Open problems.} Although multivariable MRAC has been extensively studied, it is still desirable to develop a new control scheme that enjoys a simpler controller structure than the output feedback controller and requires less feedback signals than the state feedback controller as well. In this paper, we will solve the following two problems to achieve this goal. 

\smallskip
\noindent \textbf{Problem 1. Partial-state feedback multivariable MRAC.} In this paper, we will develop and investigate a new multivariable MRAC scheme by using partial-state signal for output tracking, which can guarantee asymptotic output tracking and closed-loop system stability as well. The solution to this partial-state feedback multivariable MRAC will answer the following three technical questions:
\begin{itemize}
	\item How to use partial-state signal, instead of the state vector $x(t)$ or the output signal $y(t)$, to build up a stable MRAC scheme for achieving multivariable output tracking?
	\item How to build a unification multivariable MRAC scheme to build up a bridge between the two existing MRAC schemes? and  
	\item What are the control design flexibility and performance improvement by using partial-state feedback multivariable MRAC?
\end{itemize}

\smallskip
\noindent \textbf{Problem 2. Minimal-order multivariable MRAC.} In this paper, based on the developed partial-state feedback multivariable MRAC scheme, we will present an observer-based minimal-order multivariable MRAC scheme. The solution to this new scheme will answer the following two technical questions:
\begin{itemize}
	\item What is the least number of feedback signal for $M$-output tracking? and 
	\item How the system adaptation complexity can be reduced by the minimal-order multivariable MRAC scheme?
\end{itemize}

The study in this paper gives complete answers to the above five questions for the multivariable MRAC framework and improves the understanding of multivariable MRAC.

\smallskip
\textbf{Research background.} Besides state feedback control and output feedback control, as we mentioned above, research focusing on partial-state feedback control has been reported in the literature. In \cite{kkk95}, a partial-state feedback design is developed for nonlinear systems in a canonical form to achieve asymptotic output tracking by using a vector with a subset of state variables. In \cite{lw13b}, by using a full-order Luenberger-based state observer, an adaptive model reference controller using system measurements of dimension greater than the number of inputs is developed for bounded output tracking of multi-input-multi-output systems with $(A,B,C,C_z)$ known whose dynamics may have high relative degree and are not necessarily minimum-phase. In \cite{fk96}, the backstepping technique is utilized to construct a controller to achieve global convergence, whose design procedure may become complex when the plant order is high. In \cite{dzdh00}, link position tracking is guaranteed by a partial-state feedback controller since the requirement of the full state signal is removed by a set of filters, which is only achievable for the robotic systems under some conditions. In addition, some partial-state feedback control designs for special applications (plants) without adaptation have been developed. In \cite{u09}, partial-state feedback control is studied for a rotary crane system. In \cite{djp05}, partial-state feedback control is studied for an underactuated ship.

As we have seen, a rigorous and systematic partial-state feedback multivariable MRAC for general linear multi-input multi-output time-invariant system has not been developed yet. The new partial-state feedback multivariable MRAC scheme to be developed in this paper has less restrictive matching conditions, less state information requirement, more design flexibilities and more feedback capacities. It provides an additional and complete theoretical framework to guarantee asymptotic output of a given reference model system and closed-loop signal boundedness, in the presence of system uncertainties. It not only makes an addition to the existing family of multivariable MRAC designs and bridges the existing multivariable MRAC schemes, but also reveals some new feedback capacity of multivariable adaptive control systems. Based on the new capacity, a minimal-order MRAC system is presented. Such a new multivariable MRAC scheme requires the least number of feedback signal for multivariable output feedback control, which reduces the computation burden for control scheme implementation.

The new technical contributions of this work includes: 
\begin{itemize}
	\item developing an adaptive multivariable MRAC scheme by using partial-state feedback signal which can guarantee asymptotic output tracking and closed-loop signal boundedness in the presence of plant parameter uncertainties;
	\item conducting a complete analysis of plant-model output matching for the nominal control design, and a complete analysis of stability and tracking performance for the adaptive control design; 
	\item presenting a complete system computation complexity analysis of the partial-state feedback reduced-order multivariable MRAC scheme; and
	\item providing a minimal-order MRAC scheme which enjoys minimum feedback signal requirement and reduces the system adaptation complexity, compared to the other observer-based MRAC designs.
\end{itemize}

The rest of the paper is organized as follows. The partial-state feedback multivariable MRAC problem and the minimal-order  multivariable MRAC problem are formulated in Section 2. In Section 3, the LDS decomposition-based adaptive partial-state feedback multivariable MRAC design is developed, together with the system stability and tracking performance analysis. In Section 4, some unique features of partial-state feedback multivariable MRAC are discussed, including the inherent unification it brings to the MRAC schemes and the system complexity reduction it brings to MRAC implementation. In Section 5, the observer-based minimal-order MRAC scheme is provided. In Section 6, simulation results on an aircraft system model are presented to confirm the desired control system performance with the partial-state feedback reduced-order multivariable MRAC scheme.


\section{\textbf{Motivations and Problem Statement}}

In this section, a brief review of the existing multivariable MRAC schemes is first given in Section 2.1. Then, the multivariable MRAC problems: (a) partial-state feedback reduced-order multivariable MRAC; and (b) minimal-order multivariable MRAC, are formulated in Section 2.2.

\subsection{\textbf{Review of Multivariable MRAC Schemes}}
Before we formulate the new multivariable MRAC problems, it is necessary to review the existing multivariable MRAC schemes first in this section.

\textbf{Plant description.} Consider an $M$-input and $M$-output linear time-invariant plant described by
\begin{align}\label{ControlPlant}
\dot{x}(t) = Ax(t)+Bu(t),\; y(t) = Cx(t),
\end{align} 
where $A \in \mathbb{R}^{n \times n}$, $B \in \mathbb{R}^{n \times M}$ and $C \in \mathbb{R}^{M \times n}$ are unknown constant parameter matrices, and $x(t) \in \mathbb{R}^n$, $u(t) \in \mathbb{R}^M$ and $y(t) \in \mathbb{R}^M$ are the state, input and output vectors, respectively. The input-output description of the plant \eqref{ControlPlant} is 
\begin{equation}
y(t) = G(s)[u](t), \; G(s) = C(sI-A)^{-1}B.
\end{equation}
The notation, $y(t) = G(s)[u](t)$, is used to denote the output $y(t)$ of a system represented by a transfer function matrix $G(s)$ with a control input signal $u(t)$. It is a simple notation to combine both the time domain and the frequency domain signal operations, suitable for adaptive control system presentation.   

\textbf{Control goal and plant assumptions.} For a better understanding, we first introduce a crucial concept, as defined in the following lemma, for multivariable MRAC designs, before we describe the control goal and introduce the plant assumptions. 

\begin{lemma}\label{Definition_xi}
	\emph{\cite{t03}} For any $M \times M$ strictly proper and full rank rational matrix $G(s)$, there exists a lower triangular polynomial matrix $\xi_m(s)$, defined as the modified left interactor (MLI) matrix of $G(s)$, of the form
	\begin{align}
	\xi_m(s) = \begin{bmatrix}
	d_1(s) & 0 & \ldots & \ldots & 0 \\
	h_{21}^m(s) & d_2(s) & 0 & \ldots & 0 \\
	\vdots & \vdots & \vdots & \vdots & \vdots \\
	h_{M1}^m(s) & \ldots & \ldots & h_{MM-1}^m(s) & d_M(s) 
	\end{bmatrix},
	\end{align}
	where $h_{ij}^m(s)$, $j = 1,..., M-1, i = 2,\ldots, M$ are polynomials, and $d_i(s)$, $i = 1,..., M$ are monic stable polynomials of degrees $l_i > 0$, such that the high-frequency gain matrix of $G(s)$, defined as $K_p = \lim_{s \rightarrow \infty} \xi_m(s)G(s)$ is finite and nonsingular.
\end{lemma}

This interactor matrix $\xi_m(s)$ characterizes the plant infinity zero structure of $G(s)$, whose property of having a stable inverse is essential for MRAC designs.

The control goal of multivariable MRAC is to construct a feedback control law by using the state vector $x(t)$ or the output signal $y(t)$ for generating the control input signal $u(t)$ in \eqref{ControlPlant} such that all signals in the closed-loop system are bounded and the output vector $y(t)$ asymptotically tracks a given reference output vector $y_m(t)$ generated from a reference model system
\begin{equation}\label{ReferenceModel}
y_m(t) = W_m(s)[r](t), \, W_m(s) = \xi^{-1}_m(s),
\end{equation}
where $r(t) \in \mathbb{R}^M$ is a bounded reference input signal, and $\xi_m(s)$, defined in Lemma \ref{Definition_xi}, is a modified left interactor matrix of the system transfer matrix $G(s) = C(sI-A)^{-1}B$, whose inverse matrix is stable, i.e., $W_m(s)$ is stable.  

The basic assumptions are made for achieving the control objective for multivariable MRAC systems:
\begin{description}
	\item[(A1)] All zeros of $G(s) = C(sI-A)^{-1}B$ are stable, and $(A,B,C)$ is stabilizable and detectable.
	\item[(A2)] $G(s)$ has full rank and its modified left interactor matrix $\xi_m(s)$ is known.
\end{description}
Assumption (A1) is for a stable plant-model output matching, and Assumption (A2) is for choosing a reference model system $W_m(s) = \xi_m^{-1}(s)$ suitable for plant-model output matching. Note that the zeros of $G(s)$ are defined as the system transmission zeros (the values of $s$ making $G(s)$ nonsingular). In addition, the interactor matrix $\xi_m(s)$ does not explicitly depends on the parameters of $G(s)$ in this case. 

\smallskip
\textbf{Review of the existing MRAC designs.} According to different types of the feedback signal used to construct the controller, there are two multivariable MRAC designs for output tracking in the literature.

\smallskip
\textit{(i) State feedback for output tracking.} 
When the full state vector $x(t)$ is available for measurement, the following simple adaptive controller structure can be used:
\begin{equation}\label{StatefeedbackOutputTrackingControllerAdaptive}
	u(t) = K_1^\T(t)x(t) + K_2(t)r(t),
\end{equation}
where $K_1(t) \in \mathbb{R}^{n \times M}$ and $K_2(t) \in \mathbb{R}^{M \times M}$ are controller parameters to be adaptively updated by stable adaptive laws. Such controller parameters $K_1(t)$ and $K_2(t)$ are the adaptive estimates of the nominal controller parameters $K_1^*$ and $K_2^*$ satisfying the matching condition 
\begin{equation}\label{StatefeedbackOutputTrackingmatchingCondition}
	C(sI-A-BK_1^{*\T})^{-1}BK_2^* = W_m(s), \, K_2^{*-1} = K_p,
\end{equation}
with $K_p$ being the system high-frequency gain matrix of $G(s)$ (see Lemma \ref{Definition_xi}), for plant-model output matching: $y(t) = W_m(s)[r](t) = y_m(t)$. 
The existence of the nominal controller parameters $K_1^*$ and $K_2^*$ is guaranteed as long as the plant interactor matrix $\xi_m(s)$ is used for $W_m(s) = \xi^{-1}_m(s)$. In addition, to ensure the output tracking as well as the system internal signal boundedness, $(A,B,C)$ needs to be stabilizable and detectable and all zeros of $G(s)$ need to be stable \cite{gtl11}.

\textit{(ii) Output feedback for output tracking.}
In applications, when the full state vector $x(t)$ is not accessible for measurement, the standard output feedback adaptive controller 
\begin{align} \label{AdaptiveOutputU}
	u(t) & = \Theta_1^{\T}(t)\omega_1(t) +\Theta_2^{\T}(t)\omega_2(t) + \Theta_{20}(t)y(t) \notag \\
	& \quad + \Theta_3(t)r(t)
\end{align}
needs to be used, where $\omega_1(t) = \frac{A_0(s)}{\Lambda(s)}[u](t),\; \omega_2(t) = \frac{A_0(s)}{\Lambda(s)}[y](t)$ with $A_0(s) = [I_M,sI_M,\cdots,s^{\bar{\nu}-2}I_M]^\T$, $\Theta_1(t) \in \R^{(\bar{\nu}-1)M \times M}$, $\Theta_2(t) \in \R^{(\bar{\nu}-1)M \times M}$, $\Theta_{20}(t) \in \R^{M \times M}$, $\Theta_3(t) \in \R^{M \times M}$, $\bar{\nu}$ being the upper bound of the observability index $\nu$ of the plant, and $\Lambda(s)$ being a monic stable polynomial of degree $\bar{\nu} -1$. To ensure the internal signal boundedness while achieving output tracking, it is needed that all zeros of $G(s)$ are stable and $(A,B,C)$ needs to be stabilizable and detectable.

\textbf{Research motivations.} In summary, stable output matching can always be achieved with all zeros of $G(s)$ being stable and $(A,B,C)$ being stabilizable and detectable, and the modified left interactor matrix $\xi_m(s)$ being known as well. However, the requirement of the full state vector $x(t)$ for a state feedback controller may not be practical in applications, and the implementation complexity of an output feedback controller may also be an issue. Therefore, we develop the partial-state feedback reduced-order multivariable MRAC design, which   
\begin{itemize}
	\item increases design flexibility of multivariable MRAC systems;    
	\item introduces a unification of multivariable MRAC schemes; and
	\item provides a manageable trade-off between feedback capacity and adaptive system complexity;
\end{itemize}	
In addition, we develop the minimal-order multivariable MRAC scheme, which 
\begin{itemize}
		\item minimizes the number of feedback signal; and 
		\item minimizes the controller implementation complexity.
\end{itemize}

\begin{remark}\label{StateFeedbackStateTracking}
	\rm	
	Another MRAC system is the one which makes the plant-model state matching achievable using state feedback. The controller structure for state feedback state tracking is the same with the one for state feedback output tracking. The control objective is to make $x(t)$ track $x_m(t)$ from a chosen stable reference model system $\dot{x}_m(t) = A_mx_m(t) + B_mr(t)$. However, the plant-model matching condition: $A + BK_{1}^{*\rm T} = A_m, BK_2^* = B_m$, is restrictive for many applications, since the reference model parameters $(A_m,B_m)$ needs to be chosen in advance. In this paper, we do not consider the state tracking problem. Please refer to \cite{t03} for details. \hfill $\square$
\end{remark}

\subsection{\textbf{Partial-State Reduced-Order MRAC}}

In this paper, we will investigate the partial-state feedback reduced-order multivariable MRAC problem. Thus, besides the assumptions (A1) and (A2), we assume 
\begin{description}
	\item[(A3)] a partial-state vector signal $y_0(t) = C_0x(t) \in \mathbb{R}^{n_0}$, which is a subset of the components of $x(t)$ or a linear combination of them, is available for measurement, with $(A,C_0)$ observable for $C_0 \in \mathbb{R}^{n_0 \times n}$ and \rm{rank}$[C_0] = n_0$. 
\end{description}

\smallskip
\textbf{Problem 1. Partial-state multivariable feedback MRAC.} The control objective of this problem is to construct an adaptive control law $u(t)$ in \eqref{ControlPlant} by using the \textit{partial-state vector $y_0(t)$} such that 
\begin{itemize}
\item[] (i) all signals in the closed-loop system are bounded;
\item[] (ii) the output vector $y(t)$ asymptotically tracks the given reference output vector $y_m(t)$, i.e., $\lim_{t \rightarrow \infty}(y(t)-y_m(t)) = 0$.
\end{itemize}

\smallskip
\textbf{Problem 2. Minimal-order multivariable MRAC.} The control objective of this problem is to construct an adaptive control law $u(t)$ by using the partial-state vector \textit{$y_0(t) \in \R^{n_0}$ with a minimum $n_0$ }such that 
\begin{itemize}
	\item[] (i) all signals in the closed-loop system are bounded;
	\item[] (ii) the asymptotic $M$-output: $\lim_{t \rightarrow \infty}(y(t)-y_m(t)) = 0$, is achieved;
	\item[] (iii) the number of parameters to be adaptively updated are reduced.
\end{itemize} 

\section{New Multivariable MRAC Designs Using Partial-State Feedback}
In this section, we will first solve the partial-state feedback plant-model output matching problem by developing a new controller structure with the signal $y_0(t)= C_0x(t)$ in Section 3.1. Such a nominal controller gives the solution to the plant-model matching problem when the system parameters are known and provides a priori knowledge to the counterpart adaptive control problem which will be solved in Section 3.2. 

\subsection{\textbf{Nominal Partial-State Feedback Control}}
In this section, the nominal partial-state feedback controller is developed for the plant \eqref{ControlPlant} with known parameters, which provides the foundation for an adaptive control design with unknown parameters.

\subsubsection{\textbf{Controller Structure Development}}
In this section, we will develop a parametrized partial-state feedback controller by the partial-state $y_0(t) = C_0x(t)$ through a virtual observer. 

\textbf{Partial-State observer.} When the state $x(t)$ is not accessible, an observer-based state feedback control law: 
\begin{equation}
u(t) = K_1^{*\T}\hat{x}(t) + K_2^*r(t)
\end{equation} 
can be used for plant-model matching, with a suitable state estimate $\hat{x}(t)$. For deriving a parameterized partial-state feedback control law for plant-model output matching, we first obtain a partial-state observer with the available partial-state vector $y_0(t)$ to obtain an estimated state $\hat{x}(t)$ .

\smallskip
For the system state equation: $\dot{x}(t) = Ax(t) +Bu(t)$, as the techniques shown in \cite{c13}, we introduce a transformation matrix  $P \in \mathbb{R}^{n \times n}$ such that $C_0P^{-1} = [\begin{matrix} I_{n_0}, & 0\end{matrix}]$ with $n_0 = \textrm{rank}[C_0]$, and transfer the system state equation as  
\begin{equation}
\begin{bmatrix} \label{ReduceX}
\dot{\bar{x}}_1(t) \\ 
\dot{\bar{x}}_2(t)
\end{bmatrix} =
\begin{bmatrix}
\bar{A}_{11} & \bar{A}_{12} \\
\bar{A}_{21} & \bar{A}_{22}
\end{bmatrix}
\begin{bmatrix}
\bar{x}_1(t) \\\bar{x}_2(t)
\end{bmatrix} + 
\begin{bmatrix}
\bar{B}_1 \\\bar{B}_2
\end{bmatrix}u(t),
\end{equation}
where $\bar{x}(t) = Px(t) = [\begin{matrix}\bar{x}_1^{\T}(t), & \bar{x}_2^{\T}(t)\end{matrix}]^\T$ with $\bar{x}_1(t) \in \mathbb{R}^{n_0}$, $\bar{x}_2(t) \in \mathbb{R}^{n - n_0}$, $\bar{A}_{11} \in \mathbb{R}^{n_0 \times n_0}$, $\bar{A}_{12} \in \mathbb{R}^{n_0 \times (n-n_0)}$, $ \bar{A}_{21} \in \mathbb{R}^{(n-n_0) \times n_0}$, $\bar{A}_{22} \in \mathbb{R}^{(n-n_0) \times (n - n_0)}$, $\bar{B}_1 \in \mathbb{R}^{n_0 \times M}$ and $\bar{B}_2 \in \mathbb{R}^{(n-n_0) \times M}$. 

\smallskip
By using the techniques shown in \cite{c13}, an estimate $\hat{\bar{x}}(t)$ for $\bar{x}(t)$ can be generated as 
\begin{equation} \label{32}
\hat{\bar{x}}(t) = \begin{bmatrix} \bar{x}_1 \\ \hat{\bar{x}}_2 \end{bmatrix} = \begin{bmatrix} y_0(t) \\ w(t) + L_ry_0(t) \end{bmatrix},
\end{equation}
where $\hat{\bar{x}}_2(t)$ is an estimate for $\bar{x}_2(t)$, $L_r \in \mathbb{R}^{(n - n_0) \times n_0}$ is a constant gain matrix such that the eigenvalues of the $(n-n_0) \times (n -n_0)$ matrix $\bar{A}_{22} - L_r\bar{A}_{12}$ are stable and prespecified, and $w(t) \in \mathbb{R}^{n - n_0}$ is generated from the dynamic equation
\begin{align}\label{OmegaDifferential}
\dot{w}(t) &= (\bar{A}_{22} - L_r\bar{A}_{12})w(t) + (\bar{B}_2-L_r\bar{B}_1)u(t) \\
& \quad + ((\bar{A}_{22}-L_r\bar{A}_{12})L_r + \bar{A}_{21} - L_r\bar{A}_{11})y_0(t). \notag
\end{align}
Based on the observer-based theory, we have $\lim_{t \rightarrow \infty}(x(t) - \hat{x}(t)) = \lim_{t \rightarrow \infty}P^{-1}(\bar{x}(t) - \hat{\bar{x}}(t)) = 0$ exponentially, with the above partial-state observer.

\medskip
\textbf{Partial-state feedback controller.} The above result shows the estimate $\hat{x}(t)$ converges to $x(t)$ exponentially. Therefore, plant-model output matching should also be achievable by the observer-based control law $u(t) = K_1^{*\T}\hat{x}(t) + K_2^*r(t)$, as the nominal control law $u(t) = K_1^{*\T}x(t) + K_2^*r(t)$ does it. Since $\hat{x}(t)$ is still parameters-depending, further reparameterization of the observer-based control law is conducted for the purpose of adaptive control design for the unknown plant.
 
First, we solve the partial-state estimate $w(t)$ in (\ref{OmegaDifferential}) and express it as
\begin{align} \label{omega}
w(t) & = \varepsilon_0(t) + (sI - \bar{A}_{22} + L_r\bar{A}_{12})^{-1}(\bar{B}_2-L_r\bar{B}_1)[u](t) \notag \\
& \quad + (sI - \bar{A}_{22} + L_r\bar{A}_{12})^{-1}((\bar{A}_{22} - L_r\bar{A}_{12})L_r + \bar{A}_{21} \notag \\
& \quad - L_r\bar{A}_{11})[y_0](t) \notag \\
& = \frac{N_1(s)}{\Lambda(s)}[u](t) + \frac{N_2(s)}{\Lambda(s)}[y_0](t) + \varepsilon_0(t), 
\end{align} 
where $\varepsilon_0(t) = e^{(\bar{A}_{22}-L_r\bar{A}_{12})t}w(0)$ with $w(0)$ being an estimate of $L_ry_0(0) - \bar{x}_2(0)$, $\Lambda(s) = \textrm{det}(sI -\bar{A}_{22}+L_r\bar{A}_{12})$ whose degree is $n - n_0$ and stability properties can be prespecified by assigning the eigenvalues of $\bar{A}_{22}-L_r\bar{A}_{12}$ as a set of given (known) values, $N_1(s)$ and $N_2(s)$ are some $(n-n_0) \times M$ and $(n - n_0) \times n_0$ polynomial matrices whose maximum degrees are $n-n_0-1$ or less.

Using (\ref{32}) and (\ref{omega}), we can express the term $K_1^{*\T}\hat{x}(t)$ as 
\begin{align}\label{K1x}
K_1^{*\T}\hat{x}(t) & = \Theta_1^{*\T}\frac{A_1(s)}{\Lambda(s)}[u](t) + \Theta_2^{*\T}\frac{A_2(s)}{\Lambda(s)}[y_0](t) \notag \\
& \quad + \Theta_{20}^{*\T}y_0(t) + \varepsilon_1(t)
\end{align}
for $\varepsilon_1(t) = K_{p2}^*e^{(\bar{A}_{22}-L_r\bar{A}_{12})t}w(0)$ representing the effect of the initial condition, where $\Theta_{1}^* \in \mathbb{R}^{M(n-n_0) \times M}$, $\Theta_2^* \in \mathbb{R}^{n_0(n-n_0) \times M}$, $\Theta_{20}^* \in \mathbb{R}^{n_0 \times M}$ and $\Theta_3^* \in \mathbb{R}^{M \times M}$, such that $\Theta_{20}^{*\T} = K_{p1}^{*} + K_{p2}^{*}L_r$, $K_{p2}^{*}N_1(s) = \Theta_1^{*\T}A_1(s)$ and $K_{p2}^{*}N_2(s) = \Theta_2^{*\T}A_2(s)$, for $K_1^{*\T}P^{-1} = [K_{p1}^{*}, K_{p2}^{*}]$ with $K_{p1}^* \in \mathbb{R}^{M \times n_0}$ and $K_{p2}^* \in \mathbb{R}^{M \times (n-n_0)}$, and $A_1(s) = [I_{M}, sI_{M},\ldots,s^{n-n_0-1}I_{M}]^\T$, $A_2(s)=[I_{n_0},sI_{n_0},\ldots,$ $s^{n-{n_0}-1}I_{n_0}]^\T$.

Substituting \eqref{K1x} into the observer-based control law $u(t) = K_1^{*\T}\hat{x}(t) + K_2^*r(t)$ with $\Theta_3^* = K_2^*$ and ignoring the exponentially decaying term $\varepsilon_1(t)$, we obtain the parametrized nominal partial-state feedback controller:  
\begin{align} \label{NominalControl}
u(t) & = \Theta^{*\T}_1\omega_1(t) + \Theta_2^{*\T}\omega_2(t) + \Theta_{20}^{*\T}y_0(t) \notag \\
& \quad + \Theta_3^*r(t), 
\end{align}
where $\omega_1(t) = \frac{A_1(s)}{\Lambda(s)}[u](t), \; \omega_2(t) = \frac{A_2(s)}{\Lambda(s)}[y_0](t)$.

The above controller structure is the desired parameterized controller structure with the partial-state vector $y_0(t)$. Next, the desired plant-model output matching properties based on this controller structure are to be established.

\subsubsection{\textbf{Plant-Model Output Matching}}
The above derivation shows the partial-state feedback control law \eqref{NominalControl} is derived from the observer-based control law $u(t) = K_1^{*\T}\hat{x}(t) + K_2^*r(t)$ which is a substitution of the state feedback control law $u(t) = K_1^{*\T}x(t) + K_2^*r(t)$ when the state $x(t)$ is not available. This fact indicates that by the partial-state feedback control law \eqref{NominalControl}, desired plant-model matching can be achieved, as the other two control laws do it.


\textbf{Matching by observer-based control.} It has been shown that when $K_1^*$ and $K_2^*$ satisfy the matching condition \eqref{StatefeedbackOutputTrackingmatchingCondition}, plant-model matching can be achieved by the nominal state feedback control law: $u(t) = K_1^{*\T}x(t) + K_2^*r(t)$ \cite{gtl11}. For the same plant \eqref{ControlPlant} and the same reference model \eqref{ReferenceModel}, plant-model matching can be achieved by the nominal observer-based state feedback control law: $u(t) = K_1^{*\T}\hat{x}(t) + K_2^*r(t)$, with the same nominal parameters $K_1^*$ and $K_2^*$. The result is shown as follows.

\begin{lemma}\label{ExitenceK1K2hat}  
The observer-based state feedback controller $u(t) = K_1^{*T}\hat{x}(t) + K_2^*r(t)$, with the nominal controller parameters $K_1^{*}$ and $K_2^*$ satisfying the matching condition \eqref{StatefeedbackOutputTrackingmatchingCondition}:
\begin{equation*}
C(sI-A-BK_1^{*\T})^{-1}BK_2^* = W_m(s), \, K_2^{*-1} = K_p,
\end{equation*}
ensures plant-model output matching: $y(t) - y_m(t) = \varepsilon(t)$, for some initial condition-related exponentially decaying $\varepsilon(t)$, where $y_m(t)$ is the output of the reference model (\ref{ReferenceModel}).
\end{lemma}

\noindent\textbf{Proof:} Representing $\hat{x}(t) = x(t) + \varepsilon_2(t)$ with $\varepsilon_2(t)$ being an exponential decaying term, the observer-based control law can be expressed as $u(t) = K_1^{*\T}x(t) + K_1^{*\T}\varepsilon_2(t) + K_2^*r(t)$. Substituting this $u(t)$ into the plant \eqref{ControlPlant}, the output $y(t)$ becomes $y(t) = C(sI-A-BK_1^{*\T})^{-1}BK_1^{*\T}[\varepsilon_2](t) + C(sI-A-BK_1^{*\T})^{-1}BK_2^{*}[r](t)$. From the output matching condition: $C(sI-A-BK_1^{*\T})^{-1}BK_2^*$ $= W_m(s)$, we have $y(t) -y_m(t) = \varepsilon(t)$,
for some $\varepsilon(t)=C(sI-A-BK_1^{*\T})^{-1}BK_1^{*\T}[\varepsilon_2](t)$.  \hfill $\nabla$

Lemma \ref{ExitenceK1K2hat} confirms the existence of the nominal controller parameters $K_1^*$ and $K_2^*$ of the observer-based control law $u(t) = K_1^{*\T}\hat{x}(t) + K_2^*r(t)$, for ensuring plant-model matching.

\textbf{Matching by partial-state feedback control.}
We now present the desired output matching properties by the nominal partial-state feedback controller \eqref{NominalControl}.

\begin{theorem}	
	\it
	Constant parameters $\Theta_1^*$, $\Theta_2^*$, $\Theta_{20}^*$ and $\Theta_3^*$ exist such that the controller (\ref{NominalControl}) guarantees closed-loop signal boundedness and partial-state feedback based output matching:  $y(t) - y_m(t) = \varepsilon(t)$, for some exponentially decaying $\varepsilon(t)$.
\end{theorem}

\noindent\textbf{Proof:} The proof can be divided into two parts. The first part is for plant-model output matching by the controller \eqref{NominalControl} and the controller parameters $\Theta_1^*$, $\Theta_2^*$, $\Theta_{20}^*$ and $\Theta_3^*$, which is guaranteed based on the derivation of the partial-state feedback controller shown in Section 3.1.

The second part is for closed-loop signal boundedness. From the plant-model output matching property: $y(t) = y_m(t) + \varepsilon(t) \in L^{\infty}$, we have $\xi_m(s)[y](t) \in L^{\infty}$ since $\xi_m(s)[y_m](t) = r(t)$ and $\xi_m(s)[\varepsilon](t)$ are bounded (as $\varepsilon(t)$ is exponentially decaying).   

From $y(t) = G(s)[u](t)$ with $G(s) = C(sI-A)^{-1}B$ having full rank, ignoring the exponentially decaying effect of the initial conditions, we have $u(t) = G^{-1}(s)\xi_m^{-1}(s)\xi_m(s)[y](t)$, which is bounded, because $G^{-1}(s)\xi_m^{-1}(s)$ is stable and proper and $\xi_m(s)[y](t)$ is bounded. 

According to the full-state observer theory, for $(A,C)$ detectable, we can express the system state $x(t)$ as 
\begin{align}\label{StabilityProofObserverNominal}
x(t) & = (sI-A+LC)^{-1}B[u](t) + (sI-A+LC)^{-1}L[y](t) \notag \\
& = \frac{N_{01}(s)}{\Lambda_0(s)}[u](t) + \frac{N_{02}(s)}{\Lambda_0(s)}[y](t),
\end{align}
where the eigenvalues of the $n \times n$ matrix $A-LC$ are stable for some constant gain vector $L \in \mathbb{R}^{n \times M}$, $\Lambda_0(s) = \det(sI-A+LC)$ whose degree is $n$, $L$ is a matrix such that $N_{01}(s) = \text{adj}(sI-A+LC)B$ and $N_{02}(s)=\text{adj}(sI-A+LC)L$ are $n \times M$ polynomial matrices whose maximum degrees are $n-1$. Hence, the internal state $x(t)$ is bounded as $u(t)$ and $y(t)$ are bounded, and so is $y_0(t) = C_0x(t)$. Also, it turns out the boundedness of $\omega_1(t) = \frac{A_1(s)}{\Lambda(s)}[u](t)$, $\omega_2(t) = \frac{A_2(t)}{\Lambda(t)}[y_0](t)$. \hspace*{\fill} $\nabla$ 

Theorem 3.1 shows that when the system parameter $(A,B,C)$ are known, the partial-state feedback control law \eqref{NominalControl} with the nominal controller parameters given in \eqref{K1x} and $\Theta_3^* = K_2^*$ solves the nonadpative partial-state feedback model reference adaptive control problem.  

In addition, the nominal controller parameters $\Theta_1^*, \Theta_2^*, \Theta_{20}^*$ and $\Theta_3^*$ for output matching can also be found through a matching polynomial equation. 

\begin{corollary}
	\it
	For partial-state feedback multivariable model reference control, constant parameter matrices $\Theta_1^* \in \mathbb{R}^{M(n-n_0) \times M}$, $\Theta_2^* \in \mathbb{R}^{n_0(n-n_0) \times M}$, $\Theta_{20}^* \in \mathbb{R}^{n_0 \times M}$ and $\Theta_3^* \in \mathbb{R}^{M \times M}$ exist such that the output matching equation holds: 
	{\begin{align} \label{MatchingEquationTheorem}
		& \quad \Theta_1^{*\T}A_1(s)P(s) + (\Theta_2^{*\T}A_2(s) + \Theta_{20}^{*\T}\Lambda(s))Z_0(s) \notag \\
		& = \Lambda(s)(P(s) - \Theta_3^*K_p\xi_m(s)Z(s)).
		\end{align}}
\end{corollary}

\noindent \textbf{Proof:}
With $y_0(t) = G_0(s)[u](t)$ and $G_0(s) = C_0(sI-A)^{-1}B$, the transfer function matrix of the closed-loop system is    
\begin{align}\label{ClosedLoop1}
G_c(s) & = G(s)\big(I_M-\Theta_1^{*\T}\frac{A_1(s)}{\Lambda(s)} - (\Theta_2^{*\T}\frac{A_2(s)}{\Lambda(s)} \notag \\ 
& \quad + \Theta_{20}^{*\T})G_0(s)\big)^{-1}\Theta_3^*,  
\end{align}
which has been made to match $W_m(s) = \xi_m^{-1}(s)$. From $G_c(s) = W_m(s)$, we obtain
\begin{align}\label{MatchingEquation1}
& \quad I_M-\Theta_1^{*\T}\frac{A_1(s)}{\Lambda(s)} - (\Theta_2^{*\T}\frac{A_2(s)}{\Lambda(s)} + \Theta_{20}^{*\T})G_0(s) \notag \\
& = \Theta_3^*W_m^{-1}(s)G(s),
\end{align}
which, for $G(s) = Z(s)P^{-1}(s)$ and $G_0(s) = Z_0(s)P^{-1}(s)$, can be expressed as (\ref{MatchingEquationTheorem}).
Hence, there exist $\Theta_1^*$, $\Theta_2^*$, $\Theta_{20}^*$ and $\Theta_3^*$ satisfying the matching equation (\ref{MatchingEquationTheorem}). \hspace*{\fill} $\nabla$ 

Such a matching equation is also crucial for deriving the tracking error model for the adaptive control design in the next section.

\subsection{\textbf{Adaptive Partial-State Feedback Control}}
For the plant (\ref{ControlPlant}) with unknown $(A,B,C)$, nominal controller parameters $\Theta_1^*$, $\Theta_2^*$, $\Theta_{20}^*$ and $\Theta_3^*$ in (\ref{NominalControl}) depending on system parameters $(A,B,C)$ can not be calculated so that the nominal partial-state feedback control law cannot be applied to the plant (\ref{ControlPlant}). Thus, an adaptive partial-state feedback controller is needed to deal with the parameter uncertainties. For adaptive control, we need the following assumption:
\begin{description}
	\item[(A4)] all leading principle minors $\Delta_i$, $i = 1,2,\ldots,M$, of the high frequency matrix $K_p$ of $G(s)$, defined in Lemma \ref{Definition_xi}, are nonzero and their signs are known. 
\end{description}

\subsubsection{\textbf{Adaptive Controller and Error Model}}
In this subsection, we propose an adaptive partial-state feedback controller structure, and derive a tracking error equation.

\textbf{Controller structure.} To handle the plant (\ref{ControlPlant}) with $(A,B,C)$ unknown, we design the adaptive version of the controller \eqref{NominalControl} as
\begin{align}\label{AdaptiveController}
u(t)  & = \Theta_1^{\T}(t)\omega_1(t) + \Theta_2^{\T}(t)\omega_2(t) + \Theta_{20}^\T(t)y_0(t) \notag \\
& \quad + \Theta_3(t)r(t),
\end{align}
where $\Theta_1(t) \in \mathbb{R}^{M(n-n_0) \times M}$, $\Theta_2(t) \in \mathbb{R}^{n_0(n-n_0) \times M}$, $\Theta_{20}(t) \in \mathbb{R}^{n_0 \times M}$, $\Theta_3(t) \in \mathbb{R}^{M \times M}$ are the adaptive estimates of the unknown nominal parameters $\Theta_1^*$, $\Theta_2^*$, $\Theta_{20}^*$, $\Theta_3^*$ (defined from \eqref{K1x} or \eqref{MatchingEquationTheorem}), respectively, and 
\begin{equation}\label{OmegaInAdaptive}
\omega_1(t) = \frac{A_1(s)}{\Lambda(s)}[u](t),\; \omega_2(t) = \frac{A_2(s)}{\Lambda(s)}[y_0](t)
\end{equation}
with $A_1(s) = [I_M,sI_M,\ldots,s^{n-n_0-1}I_M]^\T$, $A_2(s) = [I_{n_0},sI_{n_0},$ $\ldots,s^{n-n_0-1}I_{n_0}]^\T$, and $\Lambda(s)$ being a monic stable polynomial of degree $n - n_0$.

\smallskip
\textbf{Estimation error model based on LDS decomposition of $K_p$.}To design an adaptive parameter update law, an error model in terms of some related parameter errors and the tracking error $e(t) = y(t) -y_m(t)$ needs to be established. 

\textit{Tracking error equation.} Recall the equation \eqref{MatchingEquation1}:
\begin{align}
& \quad I_M-\Theta_1^{*\T}\frac{A_1(s)}{\Lambda(s)} - (\Theta_2^{*\T}\frac{A_2(s)}{\Lambda(s)} + \Theta_{20}^{*\T})G_0(s) \notag \\
& = \Theta_3^*W_m^{-1}(s)G(s). \notag
\end{align} 
For $y(t) = G(s)[u](t)$ and $y_0(t) = G_0(s)[u](t)$, we operate $u(t)$ on both sides of (\ref{MatchingEquation1}), and have the signal identity:
\begin{align}\label{MatchingController}
& \quad u(t)-\Theta_1^{*\T}\frac{A_1(s)}{\Lambda(s)}[u](t) - (\Theta_2^{*\T}\frac{A_2(s)}{\Lambda(s)} + \Theta_{20}^{*\T})[y_0](t) \notag \\
& = \Theta_3^*W_m^{-1}(s)[y](t),
\end{align}
Such an equation leads to
\begin{align}\label{UForTrackingError}
u(t) & =\Theta_1^{*\T}\frac{A_1(s)}{\Lambda(s)}[u](t) +(\Theta_2^{*\T}\frac{A_2(s)}{\Lambda(s)}[y_0](t) + \Theta_{20}^{*\T}y_0(t) \notag \\
& \quad +  \Theta_3^*\xi_m(s)[y](t).
\end{align}
Substituting \eqref{AdaptiveController} from \eqref{UForTrackingError} with $r(t) = \xi_m(s)[y_m](t)$, we obtain the tracking error equation as  
\begin{equation}\label{TrackingErrorEquation}
e(t) = y(t) - y_m(t) = W_m(s)K_p[u-\Theta^{*\T}\omega](t),
\end{equation}
where $\Theta^{*} = \left[\Theta_1^{*\T}, \,\Theta_2^{*\T}, \,\Theta_{20}^{*\T}, \,\Theta_3^{*}\right]^\T$, $\omega(t) = \left[\omega_1^{\T}(t), \,\omega_2^{\T}(t) \right.$, $\left. y_0^\T(t), r^\T(t)\right]^\T$.
Such an equation can be used to develop different parameterizations for adaptive control designs, using different decompositions of $K_p$.

\textit{LDS decomposition of $K_p$.}
Given that all principle minors of the high-frequency gain matrix  $K_p$ are non-zero, the LDS decomposition of $K_p$ exists and can be employed for dealing with the uncertainty of the unknown matrix $K_p$.

\begin{lemma}
	\emph{\cite{t03}} The high-frequency gain matrix $K_p \in \mathbb{R}^{M \times M}$  with all leading principle minors nonzero has a non-unique decomposition: $K_p = L_sD_sS$, where $S \in \mathbb{R}^{M \times M}$ is such that $S = S^\T >0$, $L_s \in \mathbb{R}^{M \times M}$ is a unit upper triangle matrix, and $D_s = \diag\{s_1^*,s_2^*,\ldots,s_M^*\}$ $= \diag\{\sign[d_1^*]\gamma_1, \ldots,\sign[d_M^*]\gamma_M\}$ with arbitrary and chosen constant $\gamma_i > 0$, $i = 1,2,\ldots,M$. 
\end{lemma}

To employ this LDS decomposition of $K_p$ for adaptive control, substituting $K_p = L_sD_sS$ into the tracking error equation \eqref{TrackingErrorEquation}, with $u(t)$ from \eqref{AdaptiveController}, we have $L_s^{-1}\xi_m(s)[e](t) = D_sS\tilde{\Theta}^\T(t)\omega(t)$, where $\tilde{\Theta}(t) = \Theta(t) - \Theta^*(t)$ with $\Theta(t) = \left[\Theta_1^{\T}(t), \,\Theta_2^{\T}(t)\right.$, $\left. \Theta_{20}^{\T}(t), \,\Theta_3(t)\right]^\T$ being the estimate of $\Theta^{*} = \left[\Theta_1^{*\T}, \,\Theta_2^{*\T} \right.$, $\left. \Theta_{20}^{*\T}, \,\Theta_3^{*}\right]^\T$.

To parametrize the unknown matrix $L_s$, introducing a constant matrix $\Theta_0^* = L_s^{-1} - I = \left\{\theta_{ij}^*\right\}$ with $\theta_{ij}^* = 0$ for $i =1,2,\ldots,M$ and $j \geq i$, we have
\begin{equation}\label{trackingxi}
\xi_m(s)[e](t) + \Theta_0^*\xi_m(s)[e](t) = D_sS\tilde{\Theta}^\T(t)\omega(t).
\end{equation}
To parametrize this tracking error equation, choosing a filter $h(s) = \frac{1}{f(s)}$, where $f(s)$ is a stable and monic polynomial whose degree is equal to the maximum degree of the modified interactor matrix $\xi_m(s)$ and operating $h(s)I_M$ on both sides of \eqref{trackingxi}, we have 
\begin{align}\label{BarTrackingEquation}
& \quad \bar{e}(t) + \left[0,\,\theta_2^{*\T}\eta_2(t),\,\theta_3^{*\T}\eta_3(t),\ldots,\theta_M^{*\T}\eta_M(t)\right]^\T \notag \\
& = D_sSh(s)[\tilde{\Theta}^\T\omega](t),
\end{align}
where  
\begin{align}\label{ebar_eta_theta}
\bar{e}(t) &= \xi_m(s)h(s)[e](t) = [\bar{e}_1(t),\ldots,\bar{e}_M(t)]^\T,\notag \\
\eta_i(t) &= [\bar{e}_1(t),\ldots,\bar{e}_{i-1}(t)]^\T \in \mathbb{R}^{i-1}, i =2,\ldots,M, \notag \\
\theta_i^* &= [\theta_{i1}^*,\ldots,\theta_{ii-1}]^\T, i = 2,\ldots,M.
\end{align} 

\textit{Estimation error model.} Based on the tracking error equation \eqref{BarTrackingEquation}, we introduce the estimation error signal: 
\begin{align}\label{ErrorModel}
\epsilon(t) & = \left[0,\,\theta_2^{\T}\eta_2(t),\,\theta_3^{\T}\eta_3(t),\ldots,\theta_M^{\T}\eta_M(t)\right]^\T \notag \\
& \quad + \Psi(t)\xi(t) + \bar{e}(t),
\end{align}
with $\Psi(t)$ being the estimate of $\Psi^* = D_sS$, and 
\begin{align}\label{XiZeta}
\xi(t) &= \Theta^\T(t)\zeta(t) - h(s)[\Theta^\T\omega](t), \zeta(t) = h(s)[\omega](t).
\end{align}
It follows from \eqref{BarTrackingEquation}--\eqref{XiZeta} that
\begin{align}\label{ErrorModelWithTilde}
\epsilon(t) & = \left[0,\,\tilde{\theta}_2^{\T}\eta_2(t),\,\tilde{\theta}_3^{\T}\eta_3(t),\ldots,\tilde{\theta}_M^{\T}\eta_M(t)\right]^\T \notag \\
& \quad + D_sS\tilde{\Theta}^\T(t)\zeta(t)+ \tilde{\Psi}(t)\xi(t),
\end{align}
where $\tilde{\theta}_i(t) = \theta(t) - \theta_i^*$, $i = 2,\ldots,M$, and $\tilde{\Psi}(t) = \Psi(t) - \Psi^*(t)$ are parameter errors. Such an error equation is linear in parameter errors, which is crucial for choosing the adaptive laws for updating the controller parameters. 

\subsubsection{\textbf{Adaptive Parameter Update Law}}
Based on the error model \eqref{ErrorModelWithTilde}, the adaptive laws for updating parameter estimates are chosen as
\begin{align}
\dot{\theta}_i(t) &= -\frac{\Gamma_{\theta i}\epsilon_i(t)\eta_i(t)}{m^2(t)}, i = 2,3,\ldots,M \label{adaptivelaw1}\\
\dot{\Theta}^\T(t) &= - \frac{D_s\epsilon(t)\zeta^\T(t)}{m^2(t)},\, \dot{\Psi}(t) = -\frac{\Gamma\epsilon(t)\xi^\T(t)}{m^2(t)} \label{adaptivelaw2-3},
\end{align} 
where $\epsilon(t) = [\epsilon_1(t),\epsilon_2(t),\ldots,\epsilon_M(t)]^\T$ is computed from \eqref{ErrorModel}, $\Gamma_{\theta i} = \Gamma_{\theta i}^\T > 0$, $i = 2,3,\ldots,M$ and $\Gamma= \Gamma^\T > 0$ are adaption gain matrices, and 
\begin{equation} \label{m2}
m^2(t) = 1 + \zeta^\T(t)\zeta(t) + \xi^\T(t)\xi(t) + \sum_{i = 2}^M\eta_i^\T(t)\eta_i(t).
\end{equation}

With the positive definition function 
\begin{equation}
V = \frac{1}{2}\left(\sum_{i =2}^M\tilde{\theta}_i^\T(t)\Gamma_{\theta i}^{-1}\tilde{\theta}_i + \tr[\tilde{\Psi}^\T\Gamma^{-1}\tilde{\Psi}] + \tr[\tilde{\Theta}S\tilde{\Theta}^\T]\right),
\end{equation}
and its time-derivative $\dot{V} = \textstyle -\frac{\epsilon^\T(t)\epsilon(t)}{m^2(t)} \leq 0$, we conclude that (i) $\theta_i(t) \in L^{\infty}$, $i = 2,3,\ldots,M$, $\Theta(t) \in L^{\infty}$, $\Psi(t) \in L^{\infty}$, $\frac{\epsilon(t)}{m(t)} \in L^2 \cap L^{\infty}$, and (2) $\dot{\theta}_i(t) \in L^2 \cap L^{\infty}$, $i = 2,3,\ldots,M$, $\dot{\Theta}(t) \in L^2 \cap L^{\infty}$ and $\dot{\Psi}_i(t) \in L^2 \cap L^{\infty}$. The $L^\infty$ and $L^2$ properties of these signals are crucial for closed-loop stability, as shown next.

\subsubsection{\textbf{System Stability and Tracking Properties}}
Based on the above desired properties of the adaptive law \eqref{adaptivelaw1}--\eqref{adaptivelaw2-3}, the following desired closed-loop system properties are established.

\begin{theorem}\label{Adaptivethm}
	The adaptive partial-state feedback controller (\ref{AdaptiveController}) with the adaptive law (\ref{adaptivelaw1})--(\ref{adaptivelaw2-3}), when applied to the plant \eqref{ControlPlant}, guarantees the closed-loop signal boundedness and asymptotic output tracking: $\lim_{t \rightarrow \infty}(y(t) - y_m(t)) = 0$.
\end{theorem}

The proof of Theorem \ref{Adaptivethm} can be obtained in a similar way to that described in \cite{t03} for output feedback design. The proof is based on a well-defined feedback structure for the closed-loop system which has a small loop gain, leading to closed-loop stability. A key step in such an analysis procedure is to express a filtered version of the plant output $y(t)$ in a feedback framework which has a small gain due to the $L^2$ properties of the adaptive laws. The asymptotic tracking property follows from the complete parametrization of the error equation \eqref{ErrorModel}, the $L^2$ properties, and the signal boundedness of the closed-loop system.

To cope with the partial-state signal $y_0(t)$ in the new partial-state feedback control law, we need to express $y_0(t) = C_0x(t)$ in terms of the output $y(t)$, for which a new proof derivation is necessary. A detailed proof is shown as follows.

\textbf{Proof of Theorem \ref{Adaptivethm}.} Introduce some fictitious filters $H_i(s)$ and $K_i(s)$ as
\begin{equation}\label{FictitiousFilters}
s H_{i}(s) = 1 - K_i(s),\;K_i(s) = \frac{a_i^{d_m}}{(s + a_i)^{d_m}},\,i=1,2,3,
\end{equation}
where $a_i > 0$ is chosen to be sufficiently large and finite for $i =1,2,3$, and $d_m$ is the maximum degree of the modified interactor matrix $\xi_m(s)$ of $G(s)$.

Denote $h_i(t)$ as the impulse response functions of the transfer function $H_i(s)$, $i = 1,2,3$. From Proposition 2.10 in \cite{t03}, we have the $L^1$ operator norms 
\begin{equation}\label{H_1Impluse}
\left\| h_i(\cdot)\right\| = \frac{d_m}{a_i},\;a_i>0,\;i=1,2,3.
\end{equation} 
From $y(t) = G(s)[u](t)$ with $G(s)$ being full rank, $\omega_1(t) = F_1(s)[u](t) = \frac{A_1(s)}{\Lambda(s)}[u](t)$ in \eqref{OmegaInAdaptive} and $H_1(s)$, $K_1(s)$ in \eqref{FictitiousFilters}, we obtain
\begin{equation}\label{omega&Y-1}
F_1(s)G^{-1}(s)[y](t) = K_1^{-1}(s)[\omega_1-H_1(s)s[\omega_1]](t).
\end{equation}

To handle the new partial-state feedback control scheme, we need the transformation for the partial-state signal $y_0(t)$. First, recall the expression of internal state $x(t)$ in \eqref{StabilityProofObserverNominal}: $x(t) = \frac{N_{01}(s)}{\Lambda_0(s)}[u](t) + \frac{N_{02}(s)}{\Lambda_0(s)}[y](t)$
where $\frac{N_{01}(s)}{\Lambda_0(s)}$ and $\frac{N_{02}(s)}{\Lambda_0(s)}$ are stable and proper.

It follows that the partial-state signal $y_0(t) = C_0x(t)$ can be expressed as 
\begin{align}\label{StabilityProofY0}
y_0(t) &= C_0\frac{N_{01}(s)}{\Lambda_0(s)}[u](t) + C_0\frac{N_{02}(s)}{\Lambda_0(s)}[y](t) \notag \\ 
&=  Q_1(s)[u](t) + Q_2(s)[y](t) 
\end{align}
with $Q_1(s) = C_0\frac{N_{01}(s)}{\Lambda_0(s)}$ and $Q_2(s) = C_0\frac{N_{02}(s)}{\Lambda_0(s)}$ being stable and proper.


Let $\omega_1(s) = F_1(s)[u](t)$ have a controllable realization $(A_c,B_c)$:
\begin{equation}\label{omega1dynamic}
\dot{\omega}_1(t) = A_c\omega_1(t) + B_cu(t),
\end{equation}
where $A_c$ is a stable matrix. From \eqref{AdaptiveController}, \eqref{omega&Y-1}, \eqref{StabilityProofY0}, \eqref{omega1dynamic} and $\omega_2(t) = F_2(s)[y_0](t) = \frac{A_2(s)}{\Lambda(s)}[y_0](t)$ in \eqref{OmegaInAdaptive}  we obtain
\begin{align}\label{omega1inProof}
\omega_1(t) &= K_1(s)F_1(s)G^{-1}(s)[y](t) + H_1(s)[\dot{\omega}_1](t) \notag \\
& = K_1(s)F_1(s)G^{-1}(s)[y](t) + H_1(s)[A_c\omega_1](t)  \notag \\
& \quad + H_1(s)B_c[\Theta_1^\T\omega_1 + \Theta_2^\T F_2(s)Q_1(s)[u] \notag \\
& \quad + \Theta_2^\T F_2(s)Q_2(s)[y] + \Theta_{20}^\T Q_1(s)[u] \notag \\
& \quad + \Theta_{20}^\T Q_2(s)[y] + \Theta_3r](t).
\end{align}
Since $H_1(s)$ satisfies \eqref{H_1Impluse} and $\Theta_1(t)$ is bounded, there exists $a_1^0 >0$ such that $(I-H_1(s)(A_c+B_c\Theta_1^\T(t)))^{-1}$ is a stable and proper operator with a finite gain for any finite $a_1 > a_1^0$. The above fact can be proved similarly to the proof of Lemma 2.5 in \cite{t03}. For $0<a_1^0 <a_1$, it follows from \eqref{omega1inProof} that 
\begin{align}\label{Omega1Compact}
\omega_1(t) = G_1(s,\cdot)[u](t) + G_2(s,\cdot)[y](t) + G_3(s,\cdot)[r](t), 
\end{align}
where for $T_1(s,t) \triangleq (I-H_1(s)(A_c+B_c\Theta_1^\T(t)))^{-1}$, $G_1(s,t) = T_1(s,t)(H_1(s)B_c\Theta_2^\T(t)F_2(s)Q_1(s) + H_1(s)B_c$ $\Theta_{20}^\T(t)Q_1(s))$, $G_2(s,t) = T_1(s,t)(K_1(s)F_1(s)G^{-1}(s) + H_1(s)B_c\Theta_2^\T(t)F_2(s)Q_2(s) + H_1(s)B_c\Theta_{20}^\T(t)Q_2(s))$, $G_3(s,t) = T_1(s,t)H_1(s)B_c\Theta_3^\T(t)$ are stable and proper operators with finite gains\footnote{A linear operator $T(s, t)$ is stable and proper if $|T(s, \cdot)[x](t)| \leq \beta \int_{0}^{\T} e^{-\alpha
		(t - \tau)} |x(\tau)| d \tau + \gamma |x(t)|$ for some constants $\beta
	\geq 0$, $\alpha > 0$ and $\gamma > 0$, for all $t
	\geq 0$, where $T(s, \cdot)[x](t)$ denotes the convolution of the impulse
	response of $T(s, \cdot)$ with $x(\cdot)$ at $t$. A linear operator $T(s,t)$ is stable and strictly proper if it is stable with $\gamma = 0$.}. It follows from \eqref{Omega1Compact} with $\omega_1(t) = F_1(s)[u](t)$, $\omega_2(t) = F_2(s)[y_0](t)$ and $\omega(t) = \left[\omega_1^{\T}(t), \,\omega_2^{\T}(t) \right.$, $\left. y_0^\T(t), r^\T(t)\right]^\T$, that 
\begin{align}\label{OmegaCompact}
\omega(t) & = G_4(s,\cdot)[u](t) + G_5(s,\cdot)[y](t) \notag \\
& \quad + G_6(s,\cdot)[r](t),
\end{align}
where $G_4(s,t) = [G_1(s,t), F_1(s)Q_1(s) , Q_1(s),0]^\T$, $G_5(s,t) = [G_2(s,t), F_2(s)Q_2(s), Q_2(s),0]^\T$, $G_6(s,t) = [G_3(s,t),0,0,I]^\T$.

From \eqref{TrackingErrorEquation}, we have
\begin{equation}\label{dotYinAdaptiveProof}
\dot{y}(t) = \dot{y}_m(t) + sW_m(s)\Theta_3^{*-1}[\tilde{\Theta}^\T\omega](t). 
\end{equation}
Operating $H_2(s)$ on both sides of \eqref{dotYinAdaptiveProof} and noting that $sH_2(s) = 1-K_2(s)$, we have 
\begin{align}\label{YinAdaptiveProof}
y(t) &= K_2(s)h^{-1}(s)[\bar{y}](t) + H_2(s)sW_m(s)[r](t) \notag \\
& \quad + H_2(s)sW_m(s)\Theta_3^{*-1}\tilde{\Theta}^\T[G_4(s,\cdot)[u] \notag \\
& \quad + G_5(s,\cdot)[y] +  G_6(s,\cdot)[r]](t)
\end{align} 
with $\bar{y}(t) \triangleq h(s)[y](t)$. Similar to the operator $T_1(s,t)$, $(I - H_2(s)sW_m(s)\Theta_3^{*-1}\tilde{\Theta}^\T G_5(s,t))^{-1}$ can be proved to be a stable and proper operator with a finite gain for any finite $a_2 > a_2^0$ and some $a_2^0 > 0$. For $0 < a_2^0 < a_2$, it follows from \eqref{YinAdaptiveProof} that 
\begin{equation}\label{YCompact}
y(t) = G_7(s,\cdot)[u](t) + G_8(s,\cdot)[\bar{y}](t) + G_9(s,\cdot)[r](t),
\end{equation}
where for $T_2(s,t) \triangleq (I - H_2(s)sW_m(s)\Theta_3^{*-1}\tilde{\Theta}^\T G_5(s,t))^{-1}$, $G_7(s,t) = T_2(s,t)H_2(s)sW_m(s)\Theta_3^{*-1}\tilde{\Theta}^\T G_4(s,\cdot)$, $G_8(s,t)$ $= T_2(s,t)K_2(s)h^{-1}(s)$, 
$G_9(s,t) = T_2(s,t)H_2(s)sW_m(s)$ $(I + \Theta_3^{*-1}\tilde{\Theta}^\T G_6(s,\cdot))$ are stable and proper operators with finite gains. It follows from \eqref{OmegaCompact} and \eqref{YCompact} that 
\begin{align}\label{OmegaInProof}
\omega(t) & = (G_4(s,\cdot) + G_5(s,\cdot)G_7(s,\cdot))[u](t) \notag \\
& \quad + G_5(s,\cdot)G_8(s,\cdot)[\bar{y}](t) + (G_5(s,\cdot)G_9(s,\cdot) \notag \\
& \quad + G_6(s,\cdot))[r](t).
\end{align}
From \eqref{ErrorModel}, we express
\begin{equation}\label{Ybar}
\bar{y}(t) = \bar{y}_m(t) + W_m(s)[\epsilon - \Psi\xi - \chi](t)
\end{equation}
with $\bar{y}_m(t) = h(s)[y_m](t)$ and $\chi = \left[0,\,\theta_2^{\T}\eta_2(t),\,\theta_3^{\T}\eta_3(t) \right.$, $\left. \ldots,\theta_M^{\T}\eta_M(t)\right]^\T$. From \eqref{AdaptiveController} and \eqref{OmegaInProof}, we obtain 
\begin{align}\label{UCompact}
u(t) & = \Theta^\T(t)(G_4(s,\cdot) + G_5(s,\cdot)G_7(s,\cdot))[u](t) \notag \\
& \quad + \Theta^\T(t)G_5(s,\cdot)G_8(s,\cdot)[\bar{y}](t) \notag \\
& \quad + \Theta^\T(t)(G_6(s,\cdot) + G_5(s,\cdot)G_9(s,\cdot))[r](t).
\end{align}
From \eqref{UCompact}, it follows that 
\begin{align}\label{UCompactFinal}
u(t) &= G_{10}(s,\cdot)\Theta^\T(t)G_5(s,\cdot)G_8(s,\cdot)[\bar{y}](t)  \\
& \quad + G_{10}(s,\cdot)\Theta^\T(t)(G_6(s,\cdot) + G_5(s,\cdot)G_9(s,\cdot))[r](t) \notag 
\end{align}
where $G_{10}(s,t) = (I-\Theta^\T(t)(G_4(s,\cdot) + G_5(s,\cdot)G_7(s,\cdot)))^{-1}$ is stable and proper operators with finite gains.

From \eqref{XiZeta}, we denote $\xi(t) = [\xi_1(t),\ldots,\xi_M(t)]^\T$, $\Theta(t) = [\bar{\theta}_1^\T(t),\ldots,\bar{\theta}_M^\T(t)]^\T$ with $\bar{\theta}_i(t) \in \mathbb{R}^{(n_0+M)(n-n_0+1)}$, $i = 1,\ldots,M$ and $f(s) = s^{d_m} + \hat{a}_{d_m}s^{d_m-1} + \ldots + \hat{a}_1s + \hat{a}_0$. Then $\xi_i(t) = \bar{\theta}_i^\T(t)\zeta(t) - \frac{1}{f(s)}[\bar{\theta}_i^\T\omega](t)$, $i = 1,\ldots,M$ can be expressed as 
\begin{align}\label{Xi_i_Proof}
& \quad \xi_i(t)  \notag \\
& = \frac{s^{d_m-1} + \hat{a}_{d_m-1} s^{d_m-2} + \cdots + \hat{a}_{2} s + \hat{a}_{1}}
{f(s)}[\dot{\bar{\theta}}_i^{\T} \frac{1}{f(s)}[\omega]](t) \notag \\  
& \quad + \frac{s^{d_m-2} + \hat{a}_{d_m-1} s^{d_m-3} + \cdots + \hat{a}_{2}}
{f(s)}[\dot{\bar{\theta}}_i^{\T} \frac{s}{f(s)}[\omega]](t) \notag \\
& \quad + \cdots  + \frac{s + \hat{a}_{d_m-1}}{f(s)}[\dot{\bar{\theta}}_i^{\T}
\frac{s^{d_m-2}}{f(s)}[\omega]](t) \notag \\
& \quad + \frac{1}{f(s)}[\dot{\bar{\theta}}_i^{\T} \frac{s^{d_m-1}}{f(s)}[\omega]](t). 
\end{align}

Since $r(t) \in L^{\infty}$, from \eqref{YCompact}, \eqref{Ybar}, \eqref{UCompactFinal} and \eqref{Xi_i_Proof}, we have  
\begin{equation}\label{AbsluteZ}
\left\|\bar{y}(t) \right\| \leq x_{0} + T_3(s, \cdot)[x_{1} T_4(s, \cdot)[\left\|\bar{y}(t) \right\|](t)
\end{equation}
for some $x_{0}(t) \in L^{\infty}$, $x_{1}(t) \in L^{\infty} \cap L^{2}$ with $x_{1}(t) \geq 0$, some stable and strictly proper operator $T_{3}(s, t)$, and some stable and proper operator $T_{4}(s, t)$  with a non-negative impulse response function. It follows that
\begin{align}\label{ProofBarY}
\left\|\bar{y}(t) \right\| & \leq x_0(t) + \beta_1\int_{0}^{\T}  e^{- \alpha_1 (t - \tau)} x_{1}(\tau) \times \notag \\ 
& \quad  (\int_{0}^{\tau}  e^{- \alpha_2 (\tau - \omega)} \left\|\bar{y}(\omega) \right\|d\omega) d \tau
\end{align}
for some $\beta_1$, $\alpha_1$, $\alpha_2 > 0$.

Applying the Small Gain Lemma (Lemma 2.3 in \cite{t03}) on \eqref{ProofBarY}, we conclude that $\bar{y}(t)$ is bounded, so are $u(t)$ in \eqref{UCompactFinal} and $y(t)$ in \eqref{YCompact}. We can also obtain that $\omega(t) \in L^{\infty}$ in \eqref{OmegaInProof}, $x(t) \in L^{\infty}$ in \eqref{StabilityProofObserverNominal}, $y_0(t) \in L^{\infty}$ in \eqref{StabilityProofY0}, $\zeta(t) \in L^{\infty}$ in \eqref{XiZeta}, $\xi(t) \in L^{\infty}$ in \eqref{XiZeta}, $\bar{e}(t) \in L^{\infty}$ in \eqref{ebar_eta_theta}, $\eta_i(t) \in L^{\infty}$ in \eqref{ebar_eta_theta}, $m(t) \in L^{\infty}$ in \eqref{m2} and $\epsilon(t) \in L^{\infty}$ in \eqref{ErrorModel}. 

For $\bar{e}(t) = \xi_m(s)h(s)[e](t)$, we have 
\begin{align}\label{EinProof2}
e(t) &= W_m(s)\Theta_3^{*-1}[\tilde{\Theta}^\T\omega](t)  \\
&= H_3(s)sW_m(s)\Theta_3^{*-1}[\tilde{\Theta}^\T\omega](t) \notag \\
& \quad + W_m(s)K_3(s)h^{-1}(s)[\bar{e}](t) \notag
\end{align}
where
\begin{equation}\label{87}
\lim_{t \rightarrow \infty}W_m(s)K_3(s)h^{-1}(s)[\bar{e}](t) = 0 
\end{equation}
for a finite $a_3>0$ in $K_3(s)$, and $sW_m(s)\Theta_3^{*-1}[\tilde{\Theta}^\T\omega](t) \in L^{\infty}$. From \eqref{EinProof2} and \eqref{87}, we get
\begin{equation}\label{ProofLast}
\left\| e(t) \right\| \leq c_3\left\| h_3(t)\right\|_1 + z_1(t) \leq \frac{c_4}{a_3} + z_1(t)
\end{equation}
where $c_3$, $c_4 > 0$ and $\lim_{t \rightarrow \infty}z_1(t) = 0$. Since $a_3 > 0$ in $H_3(s)$ can be set arbitrarily large, from \eqref{ProofLast}, we can conclude that $\lim_{t \rightarrow \infty}e(t) = 0$.
\hspace*{\fill} $\nabla$

This new partial-state feedback multivariable MRAC scheme has not been reported in the literature, it has several unique features that will be discussed in the next section. 


\section{\textbf{New Features of Partial-State Feedback Multivariable MRAC Framework}} 
In this section, we discuss some advantages and unique features of the newly developed partial-state feedback adaptive control framework.
\subsection{\textbf{Unification of Multivariable MRAC}}

As we have shown so far, the use of the partial-state feedback signal $y_0(t) = C_0x(t)$ for MRAC provides new flexibilities in designing MRAC schemes. We now summarize a unified framework as follows.

\begin{proposition}
Under the standard multivariable MRAC assumptions (A1) and (A2) and the assumption (A3): $(A, C_0)$ is observable, a stable MRAC scheme using $y_0(t) = C_0x(t)$ for feedback control is capable of ensuring closed-loop signal boundedness and asymptotic output tracking, in the following cases:
	\begin{enumerate}
		\item $y_0(t)$ is a vector containing some or all elements of $y(t)$;
		\item $y_0(t)$ is vector which does not contain any element of $y(t)$;
		\item $y_0(t)$ is a scalar as one element of $y(t)$;  
		\item $y_0(t)$ is a scalar not being any element of $y(t)$; 
		\item $y_0(t)$ is the output $y(t)$; and 
		\item $y_0(t)$ is the state $x(t)$. 
	\end{enumerate}
\end{proposition}

\smallskip
Among the cases listed above, Case (3) is a special case of the Case (1) when $y_0(t) \in \R$, Case (4) is a special case of Case (2) when $y_0(t) \in \R$, Case (5) is the traditional output feedback MRAC case, Case (6) is the full-state feedback MRAC case.

The above six cases cover all kinds of possible feedback control signals available for designing a MRAC scheme to achieve asymptotic output tracking, which shows the additional design flexibility and expanded application significance of partial-state feedback MRAC. In other words, the developed partial-state feedback MRAC schemes provide a unified control framework that bridges the state feedback control MRAC and output feedback control MRAC. It is the unified solution to all multivariable MRAC problems for output tracking, and adds new design tools to the MRAC family. 

Moreover, with the use of a partial-state signal $y_0(t)=C_0x(t)$, a partial-state feedback multivariable MRAC scheme can combine the advantages of a state feedback control design and an output feedback control design. It can also provide a manageable trade-off between the two kinds of existing MRAC schemes. 

\subsection{\textbf{Reduction of Adaptation Complexity}} When $n_0$ satisfies some certain conditions, the developed partial-state feedback MRAC scheme reduces the adaptation complexity, compared to an output feedback MRAC scheme. In this paper, we use the number of updated parameters and the number of first-order integrator to measure the system adaptation complexity. 


\smallskip
\textbf{Number of updated parameters.} According to the adaptive law \eqref{adaptivelaw1}--\eqref{adaptivelaw2-3}, the total number of parameters to be updated in the partial-state feedback adaptive law \eqref{adaptivelaw1}--\eqref{adaptivelaw2-3} is 
\begin{align}
N_{ps}  & = \frac{M^2-M}{2} + (n-n_0)M^2 + (n-n_0)Mn_0 + Mn_0 \notag \\
& \quad + M^2  + M^2.
\end{align}
On the other hand, the total number of parameters to be updated in an output feedback for output tracking adaptive law is 
\begin{equation}
N_o = \frac{M^2-M}{2} + 2(\bar{\nu}-1)M^2 + 2M^2+ M^2
\end{equation}
with $\bar{\nu}$ being the upper bound of the observability index. According to \cite{t03}, the range of the observability index $\nu$ is $\frac{n}{M} \leq \nu \leq n-M+1$. Thus, we have $N_o = \frac{M^2-M}{2} + 2(n-M)M^2 + 2M^2+ M^2 = \frac{M^2-M}{2} -2M^3 + (2n+2)M^2 + M^2$. Therefore, whenever the following inequality:
\begin{align}\label{Inequality1}
& \quad N_{ps} - N_{o}  \\
& = -n_0^2 + (n+1-M)n_0 - nM-M+2M^2 < 0,\notag 
\end{align}
is satisfied, the number of parameters to be updated is reduced by the new control scheme, compared to the output feedback control scheme. By solving the inequality \eqref{Inequality1}, we conclude that for the systems with $n>3M-1$, when $n_0 < M$ or $n_0 > n-2M +1$, the number of parameters to be updated is reduced by the developed partial-state feedback scheme, and for the systems with $n<3M-1$, when $n_0 > M$, the number of parameters to be updated is reduced by the developed partial-state feedback MRAC scheme.


\smallskip
\textbf{Number of first-order integrators.} For the partial-state feedback multivariable MRAC scheme, the number of first-order integrators for constructing the filtered signals $\zeta(t)$ and $\xi$(t) is $n_h^*((M+n_0)(n-n_0+1) + M)$ with $n_h^*$ being the degree of the polynomial $f(s)$, and the number of first-order integrators for constructing $\bar{e}(t)$ is $n_e^*$ with $n_e^*$ being related to the filter $\xi_m(s)h(s)$. Therefore, the total first-order integrators used for partial-state feedback control adaptation is $N^{\prime}_{ps} = n_h^*((M+n_0)(n-n_0+1)+M)+ n_e^*$. Similarly, the number of first-order integrators used for output feedback control adaptation is $N^{\prime}_o = n_h^*(2\bar{\nu} M + M) + n_e^*$ with $\bar{\nu} = n-M+1$. Therefore, whenever the following inequality:
\begin{align}\label{Inequality2}
& \quad N^{\prime}_{ps} - N^{\prime}_{o}  \\
& = n_h^*(-n_0^2 + (n+1-M)n_0 -nM - M + 2M^2) < 0, \notag 
\end{align}
is satisfied, the number of first-order integrators used for control adaptation is reduced by the partial-state control scheme. By solving the inequality \eqref{Inequality2}, we conclude that for the systems with $n>3M-1$, when $n_0 < M$ or $n_0 > n-2M +1$, the number of first-order filters is reduced by the developed partial-state feedback scheme, and for the systems with $n<3M-1$, when $n_0 > M$, the number of first-order filters is reduced by the developed partial-state feedback scheme. 

\smallskip
Summarizing the above results, we can make the following conclusion. 

\begin{proposition}
		For a plant in the form of \eqref{ControlPlant} with $n>3M-1$, the adaptation complexity is reduced by the partial-state feedback multivariable MRAC scheme using the partial-state $y_0(t) \in \R^{n_0}$ with the condition $n_0 < M$ or $n_0 > n-2M +1$; For a plant in the form of \eqref{ControlPlant} with $n<3M-1$, the adaptation complexity is reduced by the partial-state feedback multivariable MRAC scheme using the partial-state $y_0(t) \in \R^{n_0}$ with the condition $n_0 > M$. 
\end{proposition}

\section{\textbf{Toward Minimal-Order Multivariable MRAC}}

In this section, we will present an observer-based minimal-order multivariable MRAC scheme, which allows the least number of feedback signals for multivariable feedback control and significantly reduces the system complexity compared to an output feedback control scheme.

\subsection{\textbf{MRAC with Minimum Feedback Signals}} Recall the six possible cases (options) of using the feedback signal $y_0(t) = C_0 x(t)$ for MRAC design, as summarized in Section 4.1, in particular, Case (3): $y_0(t) \in R$ is a scalar component of the system output vector $y(t) \in R^M$, and Case (4): $y_0(t) \in R$ is a scalar signal not a component of $y(t) \in R^M$. For these two cases, from the developed partial-state feedback MRAC scheme, we have the controller structure as follows:
\begin{align}\label{MinimumOrderController}
u(t)  & = \Theta_1^{\T}(t)\omega_1(t) + \Theta_2^{\T}(t)\omega_2(t) + \Theta_{20}^\T(t)y_0(t)\notag \\
& \quad + \Theta_3(t)r(t),
\end{align}
where $\Theta_1(t) \in \R^{M(n-1) \times M}$, $\Theta_2(t) \in \R^{(n-1) \times M}$, $\Theta_{20}(t) \in \R^{1 \times M}$, $\Theta_3(t) \in \R^{M \times M}$ are the adaptive estimates of the unknown nominal parameters $\Theta_1^*$, $\Theta_2^*$, $\Theta_{20}^*$, $\Theta_3^*$ (defined from \eqref{K1x}), respectively, and $\omega_1(t)$, $\omega_2(t)$ are in the form of \eqref{OmegaInAdaptive} with $A_1(s) =
[I_M,sI_M,\ldots,s^{n-2}I_M]^\T$, $A_2(s) = [1,s,\ldots,s^{n-2}]^\T$,
and $\Lambda(s)$ being a monic stable polynomial of degree $n - 1$.

From Theorem \ref{Adaptivethm}, the controller structure \eqref{MinimumOrderController}, with the feedback signal $y_0(t)$ being a scalar, guarantees $M$-output tracking for a multivariable plant. Such an adaptive control scheme, when applied to Case (3), shows that it is sufficient for the controller to only use one component of $y(t)$ for feedback control to achieve $M$-output tracking; and when applied to Case (4), it shows that the controller can only use a scalar signal $y_0(t)$ (which is not even from the components of $y(t)$) for feedback control to achieve $M$-output tracking. Such a result has not been reported in the literature and is believed to be a novel concept in adaptive control.  We formally summarize it as follows. 

\begin{proposition}\label{AdaptiveCor} 
	A partial-state feedback multivariable MRAC scheme can be designed, only using a scalar signal $y_0(t) = C_0 x(t) \in R$ with $(A,C_0)$ observable for feedback, to construct the adaptive feedback controller \eqref{MinimumOrderController} for a multi-input multi-output system: $\dot{x}(t) = Ax(t)+Bu(t)$, $y(t) = Cx(t) \in R^M$, to achieve the desired system properties: closed-loop signal boundedness and asymptotic output tracking: $\lim_{t \rightarrow \infty}(y(t) - y_m(t)) = 0$ for a given reference output signal $y_m(t) \in \mathbb{R}^M$. 
\end{proposition}

The controller \eqref{MinimumOrderController} only uses a scalar signal $y_0(t) \in \R$ from the controlled system for feedback control design, to be able to guarantee an $M$-output tracking, which minimizes the amount of feedback signals for constructing the adaptive controller.

\subsection{\textbf{Reduction of Adaptation Complexity}} 
From the review in Section 1, it is concluded that the high-order of an output feedback controller confines its application although it does not require full-state vector information. In this section, we will show that the minimal-order controller \eqref{MinimumOrderController}, without the requirement of the full-state vector, reduces the order of control to cover more control applications. 

Substituting the condition $n_0 =1$ into the inequality \eqref{Inequality1} and \eqref{Inequality2}, we obtain an equivalent inequality: $n-M-nM-M+2M^2<0$, for finding the condition that makes the adaptation complexity of the minimal-order multivariable control less than the one of the output feedback multivariable control. Solving this inequality, we can readily conclude that the inequalities hold when $M < \frac{n}{2}$. Such a result means that for MIMO systems ($M \geq 2$), the system adaptation complexity (i.e., the number of updated control parameters and the number of first-order integrator) are reduced by the controller structure \eqref{MinimumOrderController}, when the output dimension $M$ is less than the half of the state dimension $n$. Such an adaptation complexity reduction condition is often the case of real multivariable control systems, such as the aircraft control system shown in Section \ref{SimulationStudy}. 

In addition, we can also conclude that when $M = \frac{n+2}{4}$, the function $f(M) = n-M-nM-M+2M^2$ has the minimal value: $f(M) = -\frac{1}{8}(n-2)^2$. In other words, compared to the output feedback output tracking scheme, the number of parameters to be adaptively updated and the number of first-order integrators used in the adaptive control system can be reduced up to $\frac{1}{8}(n-2)^2$, when the output dimension $M$ is chosen as $\frac{n+2}{4}$, by using the minimal-order multivariable control scheme. Such a result is also helpful for the choice of system output.

\smallskip
\begin{proposition}\label{prop_Mini}
	For multivariable model reference adaptive control systems, as long as the dimension $M$ of the plant output $y(t)$ is less than $\frac{n}{2}$, the system adaptation complexity is reduced by the minimal-order controller \eqref{MinimumOrderController}. In particular, for the control system with $y_0(t) \in \R$, when $M = \frac{n+2}{4}$, the system adaptation complexity is minimized, which is $\frac{1}{8}(n-2)^2$ less than the output feedback multivariable MRAC system.
\end{proposition}

So far, we have confirmed the two features of the minimal-order multivariable controller \eqref{MinimumOrderController}: (a) the amount of feedback signal used for constructing the feedback controller is minimum; and (b) the system adaptation complexity can be reduced. 

\noindent\makebox[\linewidth]{\rule{\paperwidth}{0.4pt}}
\begin{figure*}[htb!]
	\small
	\rule{\linewidth}{0.4pt}
\begin{align} \label{SimulationParamter}
A & = \begin{bmatrix}
-0.019 & 0.1364 & -9.7778 & -32.0829 &  -0.0018 & -0.0004 & 0 & 0 \\
-0.2804 & -2.7567 & 120.1968 & -2.42 &  -0.0001 & 0 & 0.0004 & -0.0061\\
0.0205 & -0.3106 & -3.5393 & 0 &  0.007 & 0.0328 & -0.0014 & 0 \\
0 & 0 & 1 & 0 &       0   & -0.0002 & 0 & 0.0002 \\
0 & -0.0027 & 0 &     -0.0005     &  -0.5765 & -125.9974 & 10.4690 & 32.0829 \\
0 & 0 & -0.0255 &     0     &     0.2245     & -1.4053 & -0.2794 & 0\\
0 & 0 & 0.0018 & 0 &  -0.629 & 1.9689 & -5.4759 & 0 \\
0 & 0 & 0 & -0.0002 &  0 & 0.0754 & 1 & 0
\end{bmatrix},  \notag \\
B & = \begin{bmatrix}
0.0056 & -0.0423 \\
-0.6119 & 0.1579 \\
-0.7486 & 0.0859 \\
0 & 0 \\
0 & -0.0223 \\
0 & -0.0223 \\
0 & -0.7657 \\
0 & 0 
\end{bmatrix}, \;\;\;\;\; 
C  = \begin{bmatrix}
0 & 0 & 0 & 1 & 0 & 0 & 0 \\
0 & 0 & 0 & 0 & 0 & 0 & 1 
\end{bmatrix}.
\end{align}
	\rule{\linewidth}{0.4pt}
\end{figure*}

\section{Simulation Study}\label{SimulationStudy}
In this section, we present a simulation study to evaluate the effectiveness of the proposed partial-state feedback adaptive control designs.
 
\subsection{\textbf{Simulation System}}
The NASA GTM model \cite{ltj10} is chosen as the plant, which the proposed partial-state feedback adaptive control design is applied on.  

\textbf{Plant dynamics.} The linearized NASA GTM model is in the form of \eqref{ControlPlant}: $\dot{x} = Ax+Bu,y=Cx$. The system state vector is $x = [u_b,w_b,q_b,\theta,v_b,r_b,p_b,\phi]^\T$ with $u_b$, $v_b$, $w_b$ being the body-axis velocity components of origin of body-axis frame, $p_b$, $q_b$ and $r_b$ being the body-axis components of angular velocity and $\theta$, $\phi$ being the pitch and roll angle. The control inputs are the elevator angular $\delta_e$ and the aileron angular $\delta_a$, and the plant outputs are chosen as the pitch angle $\theta$ and the roll angle $\phi$. The system parameter matrices are shown in \eqref{SimulationParamter}. 

\textbf{Verification of design conditions.} For the aircraft model $(A,B,C)$ in \eqref{SimulationParamter}, it can be verified that the transfer function $G(s) = C(sI-A)^{-1}B$ has stable zeros: $s_{1,2} = -1.0059\pm5.5340i$, $s_3 = -2.4867$ and $s_4 = -0.035$, and $G(s)$ is strictly proper and full rank. The modified interactor matrix $\xi_m(s)$ can be chosen as $\xi_m(s) = \diag \left\{(s+2)^2, (s+2)^2\right\}$ so that
\begin{align}\label{KpinSimulation}
K_p = \lim_{s \rightarrow \infty}\xi_m(s)G(s) = \begin{bmatrix}
-0.7486 & 0.0859 \\ 
-0.00001 & -0.7675
\end{bmatrix}
\end{align}
is finite and non-singular. From \eqref{KpinSimulation}, the design condition that the signs of leading principle minors of $K_p$ are positive can also be verified. 

\textbf{Reference model.} Since the modified interactor matrix $\xi_m(s)$ is chosen as $\diag \left\{(s+2)^2, (s+2)^2\right\}$, the transfer function of the chosen reference model \eqref{ReferenceModel} is 
\begin{equation}
W_m(s) = \xi_m^{-1}(s) = \diag \left\{\frac{1}{(s+2)^2}, \frac{1}{(s+2)^2}\right\},
\end{equation}
which is proper and stable. The reference inputs are chosen as $r(t) = [-40\pi/180\sin(0.1t)\,-15\pi/180\sin(0.1t)]^\T$.

\begin{figure}	
	\centering
	\includegraphics[width=\linewidth]{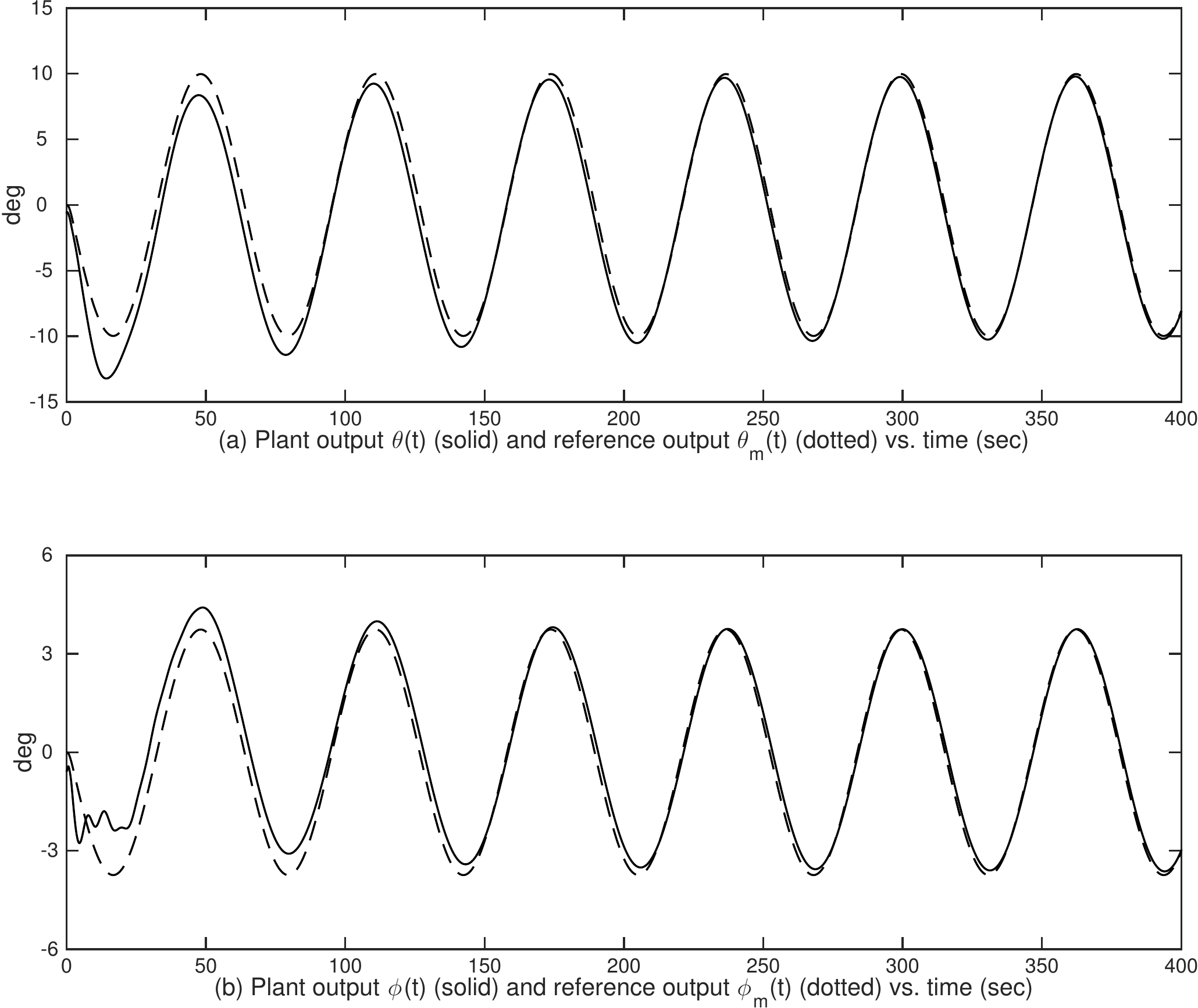}
	\caption{Plant Output: pitch angle $\theta$ and roll angle $\phi$ in Case I.}
	\label{fig:Case1_Output}
	\centering
	\includegraphics[width=\linewidth]{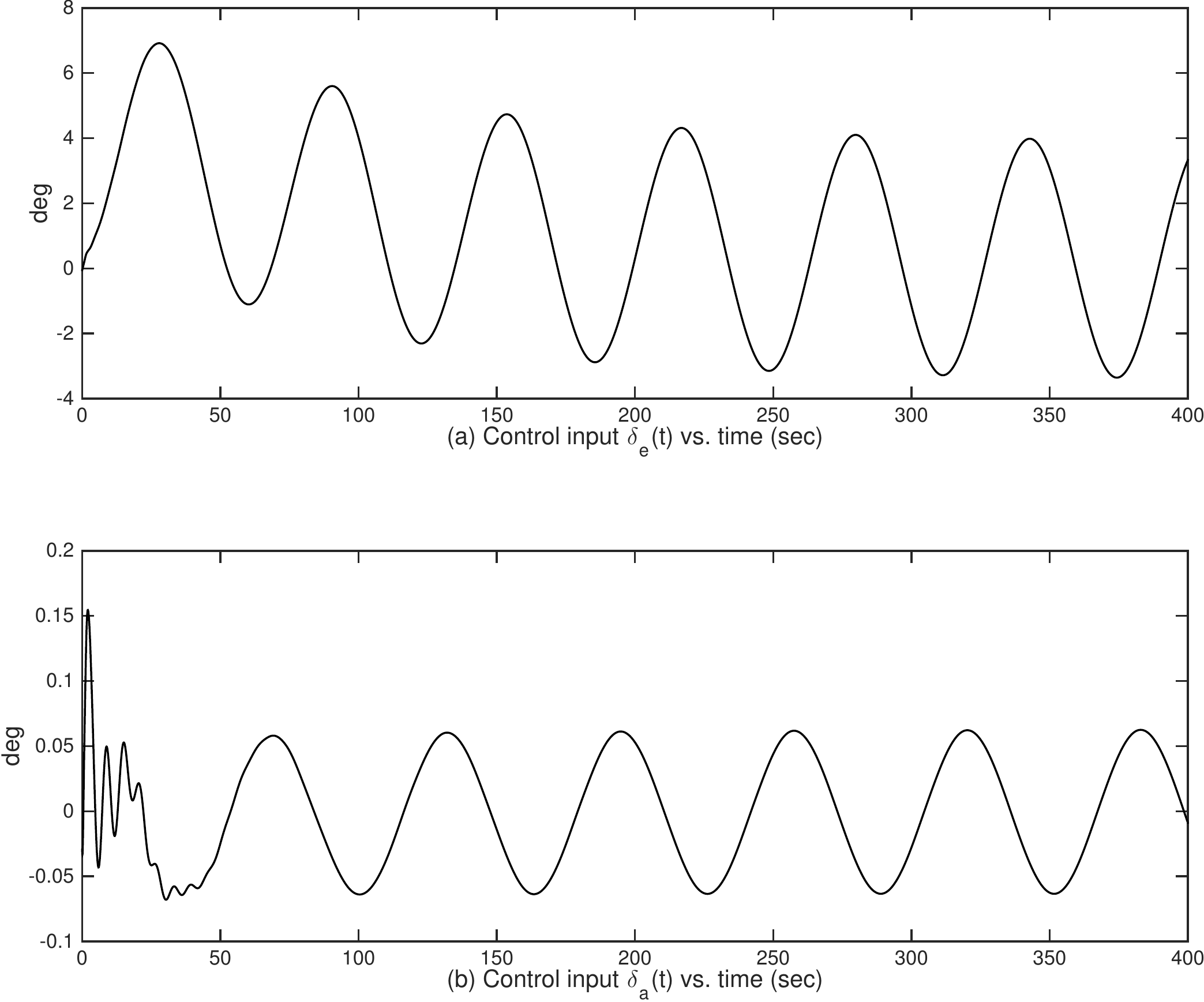}
	\caption{Control input signal: elevator angle $\delta_e$ and aileron angle $\delta_a$ in Case I.}
	\label{fig:Case1_Input}
\end{figure}

\begin{figure}
	\centering
	\includegraphics[width=\linewidth]{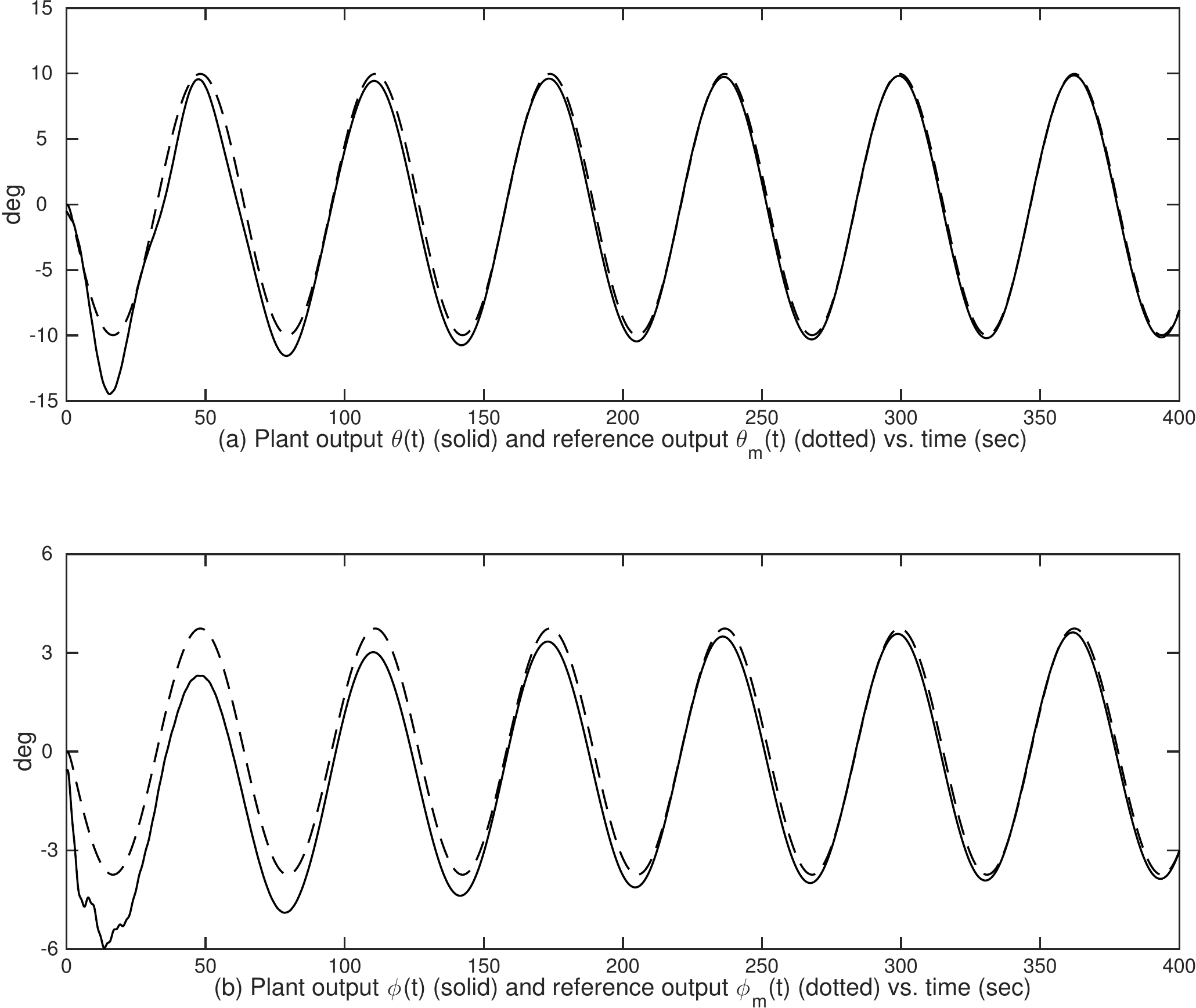}
	\caption{Plant output: pitch angle $\theta$ and roll angle $\phi$ in Case II.}
	\label{fig:Case2_Output}
\end{figure}
\begin{figure}
	\centering
	\includegraphics[width=\linewidth]{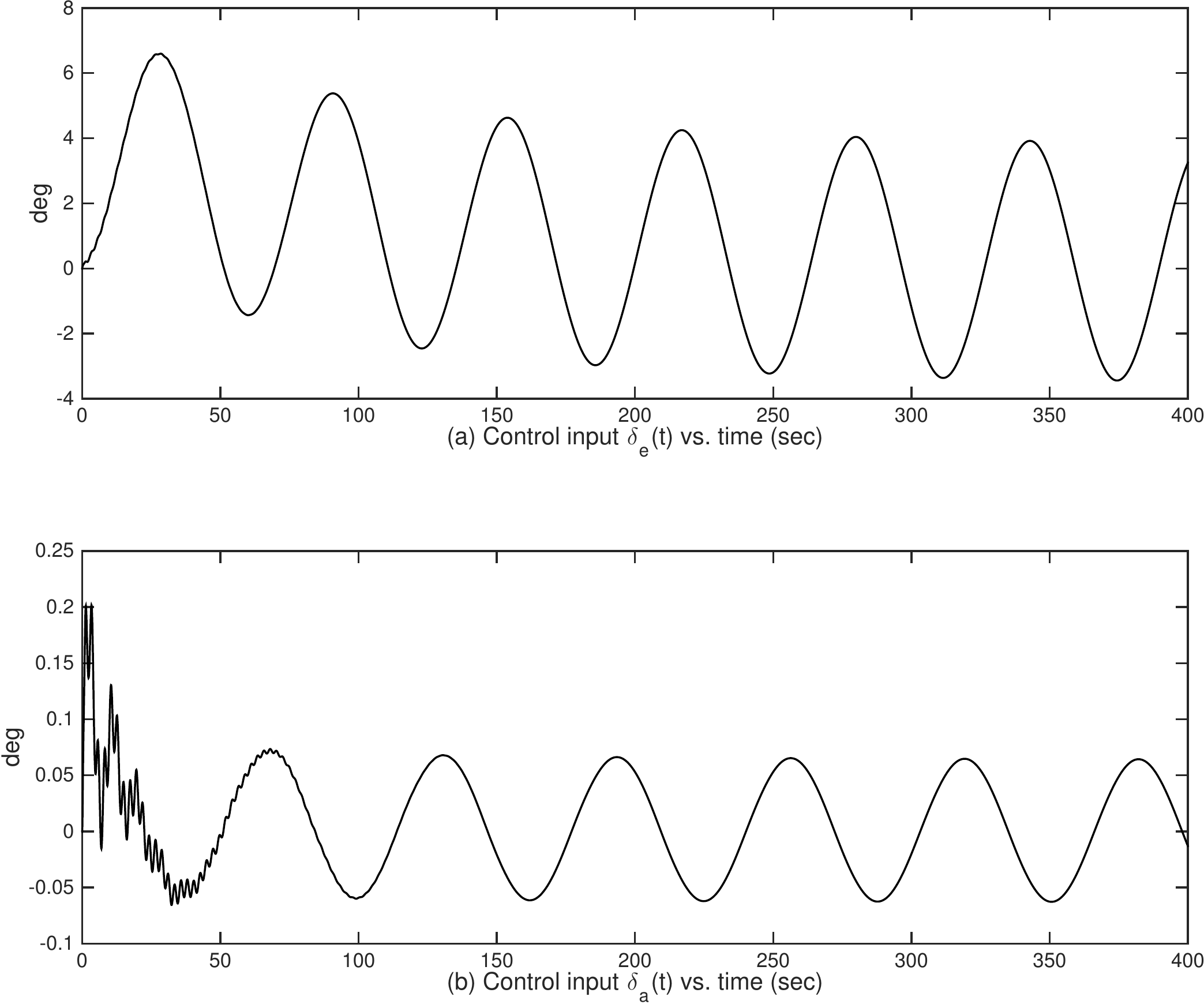}
	\caption{Control input signal: elevator angle $\delta_e$ and aileron angle $\delta_a$ in Case II.}
	\label{fig:Case2_Input}
\end{figure}

\begin{figure}
	\centering
	\includegraphics[width=\linewidth]{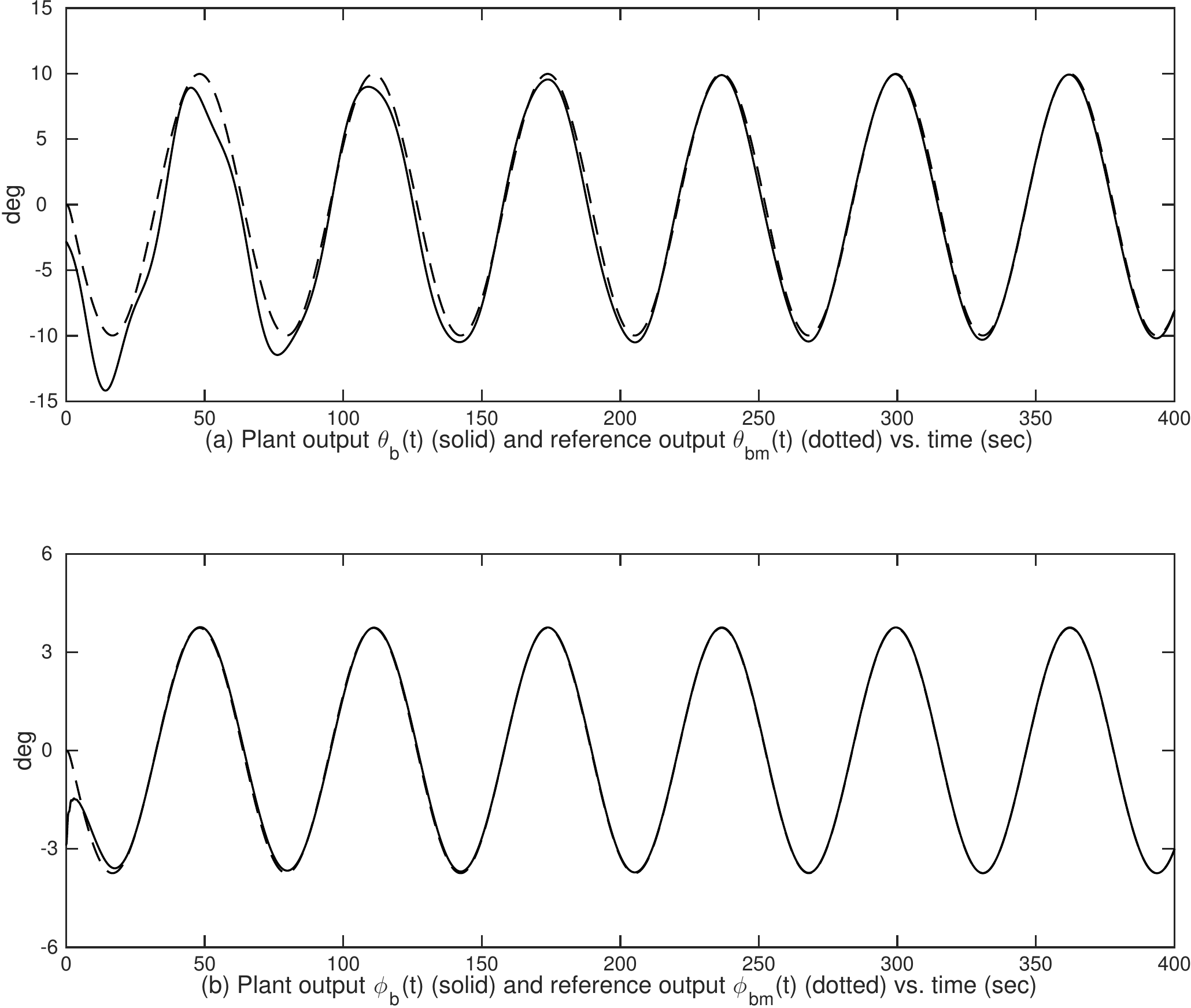}
	\caption{Plant output: pitch angle $\theta$ and roll angle $\phi$ in Case III.}
	\label{fig:Case3_Output}
\end{figure}
\begin{figure}
	\centering
	\includegraphics[width=\linewidth]{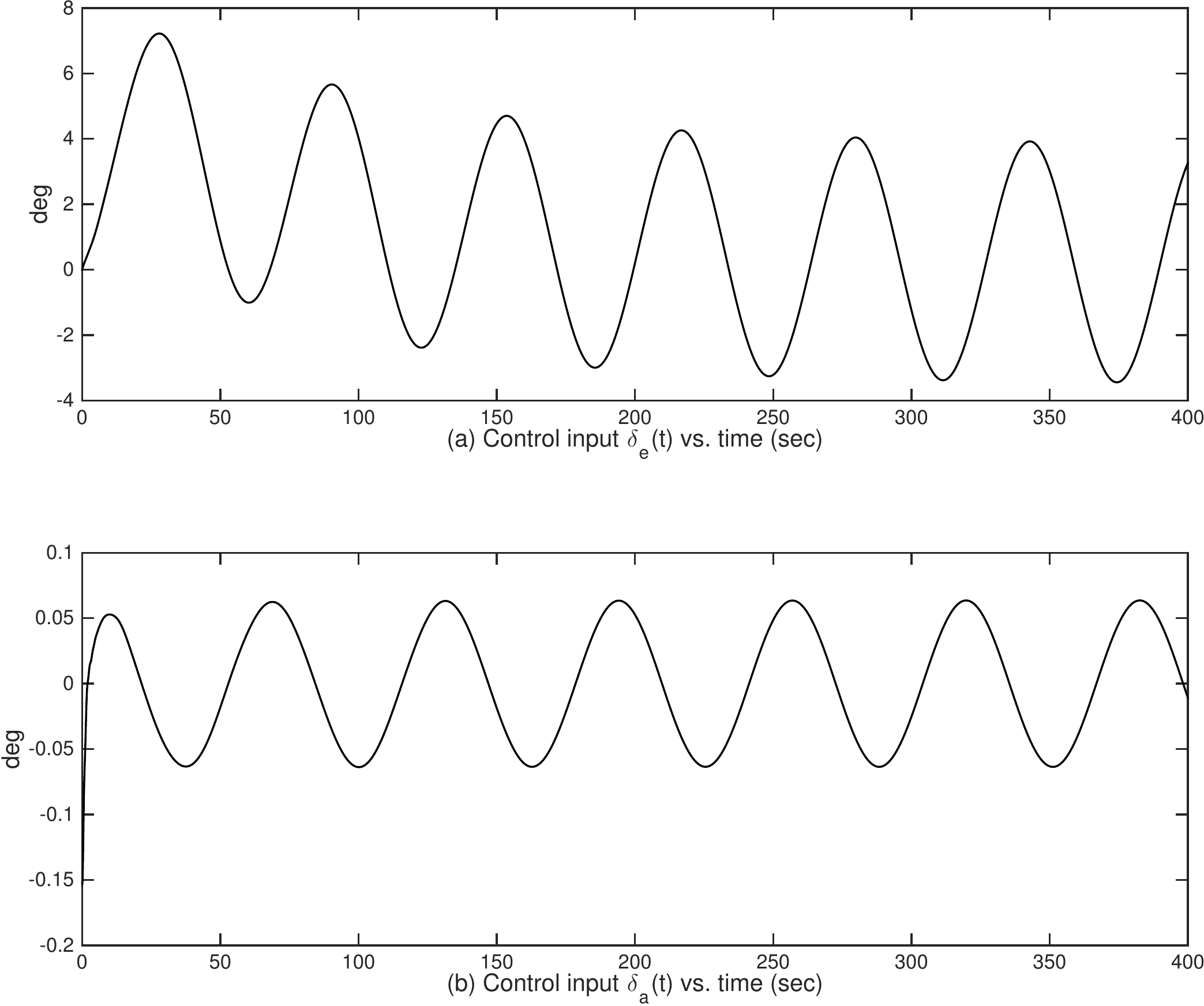}
	\caption{Control input signal: elevator angle $\delta_e$ and aileron angle $\delta_a$ in Case III.}
	\label{fig:Case3_Input}
\end{figure}

\begin{figure}
	\centering
	\includegraphics[width=\linewidth]{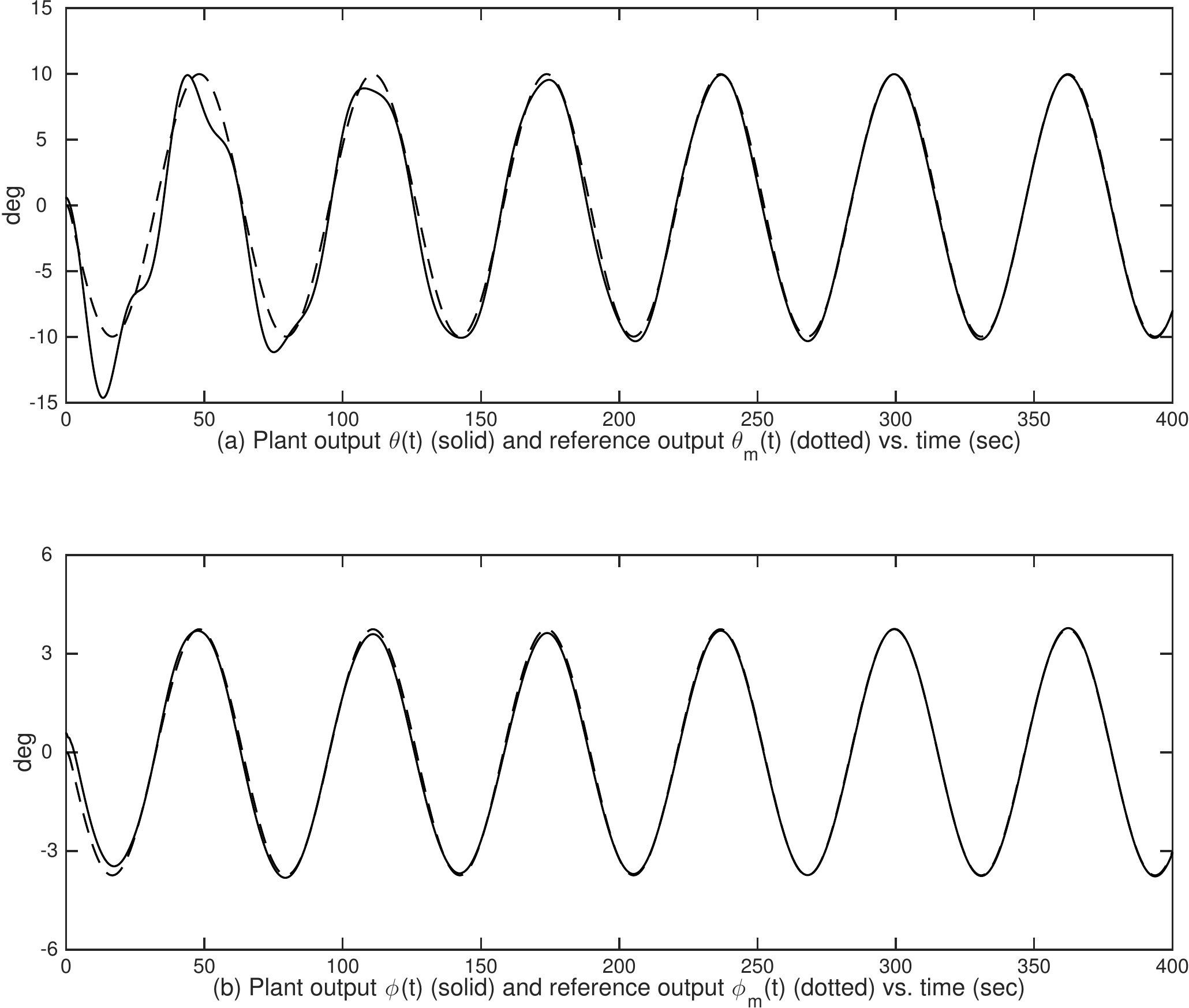}
	\caption{Plant output: pitch angle $\theta$ and roll angle $\phi$ in Case IV.}
	\label{fig:Case4_Output}
\end{figure}
\begin{figure}
	\centering
	\includegraphics[width=\linewidth]{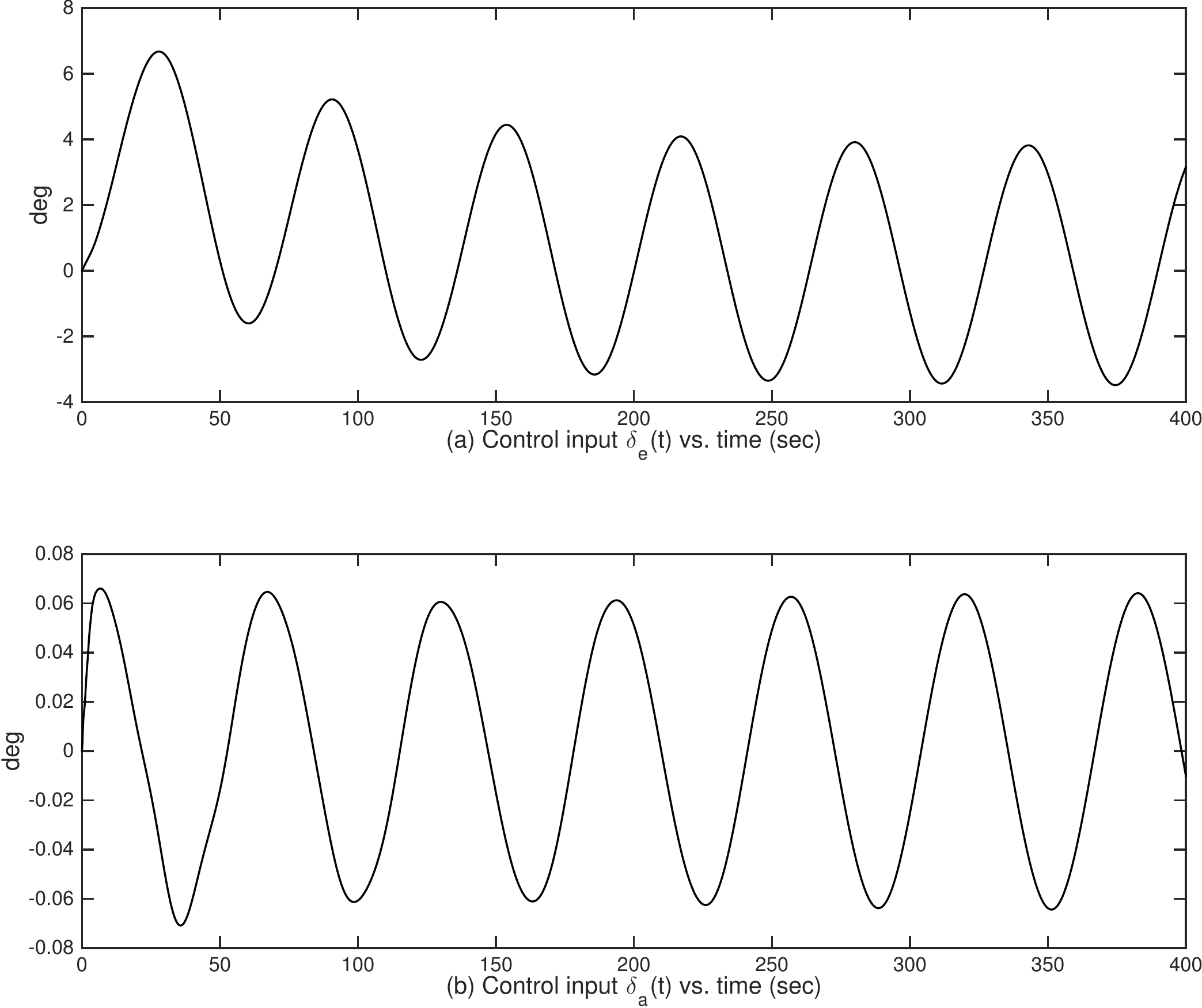}
	\caption{Control input signal: elevator angle $\delta_e$ and aileron angle $\delta_a$ in Case IV.}
	\label{fig:Case4_Input}
\end{figure}

\subsection{\textbf{Simulation Results}}
Four cases have been systematically studied to show the new features of the partial-state feedback MRAC scheme.
\begin{description}
\item[] \textbf{Case I:} $y_0(t) = [q_b,\theta,p_b]^\T$ is a vector containing one element of $y = [\theta,\phi]^\T$;
\item[] \textbf{Case II:} $y_0(t) = [q_b,r_b,p_b]^\T$ is vector which does not contain any element of $y = [\theta,\phi]^\T$;
\item[] \textbf{Case III:} $y_0(t) = \phi(t)$ is a scalar as one element of $y(t) = [\theta,\phi]^\T$; and 
\item[] \textbf{Case IV:} $y_0(t) = r_b(t)$ is a scalar not being any element of $y(t)= [\theta,\phi]^\T$. 
\end{description}

For all simulation cases, the adaptation gains are chosen as $\Gamma = 5I$, $\Gamma_{\theta} = 5$, and the initial condition are chosen as $y(0) =[-0.01,-0.01]^{\T}$, $y_m(0) = [0,0]^{\T}$. Case I and II tests the plant output tracking performance when the partial-state feedback signal $y_0(t)$ are vectors, and Case III and IV tests the tracking performance when the partial-state feedback signal $y_0(t)$ are scalars. The plant output tracking performances of Case I -- Case IV are shown in Fig. \ref{fig:Case1_Output}, Fig. \ref{fig:Case2_Output}, Fig. \ref{fig:Case3_Output} and Fig. \ref{fig:Case4_Output}, respectively, in which the dotted lines represent the reference pitch angle and roll angle and the solid lines represent the aircraft outputs. The tracking performance plots show that the asymptotic tracking are achieved in all four cases, in particular, the one for Case III and IV confirms the result in Corollary \ref{AdaptiveCor} that, for partial-state feedback MRAC, a scalar feedback signal is sufficient for constructing an adaptive controller to make the $M$ output tracking achievable.

Also, for Case III and IV, adaptive controllers are constructed based on \eqref{MinimumOrderController} whose parameter order is 48. While for the same plant, if a standard output feedback controller \eqref{AdaptiveOutputU} is constructed, the controller parameter order will be 56, since the upper bound of the plant observability index $\bar{\nu}$ is 7, which supports that results in Proposition \ref{prop_Mini}.

Moreover, Fig. \ref{fig:Case1_Input}, Fig. \ref{fig:Case2_Input}, Fig. \ref{fig:Case3_Input} and Fig. \ref{fig:Case4_Input} show the control input signals of Case I -- Case IV, respectively, which confirm that all control signals stay in acceptable ranges. In addition, signals in closed-loop systems for all four cases are bounded whose plots are not shown due to the space limit.

\section{Conclusions}
In this paper, we have developed a new framework of multivariable MRAC using partial-state feedback for output tracking, with new solutions to three technical issues: plant-model output matching, parameterized error model based on LDS decomposition, and stable adaptive law design and analysis, for ensuring closed-loop system stability and asymptotic tracking in the presence of plant uncertainties. This work has shown that partial-state feedback MRAC provides additional design flexibilities in utilizing system signals, while using less complex controller structures than output feedback. We presented a complete analysis of the closed-loop system stability and tracking performance of partial-state feedback MRAC. It has been shown that such a new MRAC framework builds a natural transition from full state feedback MRAC to output feedback MRAC, adding new members to the family of MRAC. Moreover, we conclude that for the partial-state feedback MRAC scheme, asymptotic tracking for $M$ ($M \geq 1$) output is achievable by the adaptive controller constructed by some scalar feedback signals, and provide an observer-based minimal-order MRAC scheme based on which. We presented simulation results for different adaptive control designs, which verify the desired adaptive control system performance.



\end{document}